\documentclass[manuscript]{acmart}
\usepackage{tikz}
\usetikzlibrary{positioning}
\usetikzlibrary{arrows}
\usetikzlibrary{shapes}
\usepackage{longtable,booktabs}
\usepackage{array}
\usepackage{hyperref}
\usepackage[table]{xcolor} 
\usepackage{collcell}       
\usepackage{pgf}            
\usepackage{draftwatermark}

\SetWatermarkText{Submitted to ACM CSUR}
\SetWatermarkScale{0.3}
\SetWatermarkColor[gray]{0.85} 

\newcounter{minval}
\newcounter{maxval}
\newcommand{\ApplyGradient}[1]{%
    \pgfmathsetmacro{\PercentColor}{%
        100 * ((#1 - \value{minval}) / (\value{maxval} - \value{minval}))
    }%
    \pgfmathsetmacro{\ClampedColor}{max(0, min(100, \PercentColor))}%
    \cellcolor{blue!\ClampedColor!white}#1%
}

\newcolumntype{G}{>{\collectcell\ApplyGradient}c<{\endcollectcell}}

\AtBeginDocument{%
  }

\setcopyright{acmlicensed}
\copyrightyear{2025}
\acmYear{2025}
\acmDOI{XXXXXXX.XXXXXXX}

\begin{document}


\title[Converging Zero Trust and IoT Security: A MLR]{Converging Zero Trust and IoT Security: A Multivocal Literature Review}

\author{Mariam Wehbe}
\email{mariam.wehbe@insa-cvl.fr}
\orcid{0009-0009-0241-7284}
\author{Laurent Bobelin}
\email{laurent.bobelin@insa-cvl.fr}
\orcid{0000-0002-3268-4203}
\affiliation{%
  \institution{INSA Centre Val de Loire}
  \city{Bourges}
  \country{France}
}



\begin{abstract}
The convergence of Internet of Things (IoT) security and Zero Trust (ZT) principles is a trending topic, demanding a comprehensive, multi-perspective analysis. We present the first multivocal literature review (MLR) on this topic, combining 68 academic and 36 industrial studies. This comprehensive review identifies two complementary yet divergent perspectives: academia focuses on IoT compliance with ZT principles through IoT modifications, while industry prioritizes practical integration within existing ZT frameworks guided by NIST standards. The analysis reveals critical research gaps in socio-technical understanding, cost-benefit evaluation, and interdisciplinary collaboration, highlighting these as key directions for future research.

\end{abstract}



\received{}
\received[revised]{}
\received[accepted]{}

\maketitle

\section{Introduction}
Zero Trust (ZT) security \cite{NIST} was designed to avoid implicit trust - a philosophy summarized by the motto \textit{``Never trust, always verify"}. Since Google’s seminal BeyondCorp initiative \cite{beyondcorp}, both academia and industry have shown growing interest in ZT. For the former, it proposes exciting new challenges, for the latter, ZT enforces security at an unprecedented level. 

 Internet of Things (IoT) devices are now widely deployed around the world. 
 Modern systems commonly combine IoT devices with server, cloud, fog, and edge-based computation and storage, as seen in IIoT, smart systems, and dataspaces. 
 IoT security is considered a specific subdomain of cybersecurity \cite{Rajmohan2022}. Integrating IoT devices increases the complexity of adopting ZT: IoT and ZT security paradigms must converge to form a consistent model. This issue has been examined by numerous researchers and industry stakeholders. However, despite the growing adoption of ZT, no comprehensive synthesis exists of how ZT principles apply to IoT security in both academic and industrial contexts.

 
The convergence between ZT and IoT security can be approached in two ways: (1) align IoT security with ZT principles to create a ZT-compliant model for IoT platforms, or (2) integrate IoT security into a broader ZT-based architecture.   

Many challenges exist for the first approach — making IoT systems follow ZT guidelines. ZT requires end-to-end security but IoT is a vulnerable endpoint. IoT devices may lack the capacity to encrypt traffic or authenticate themselves, yet encryption and authentication are core ZT requirements.Even when IoT devices implement these features, they may still lack the capacity to protect credentials or apply strong encryption. As ZT rejects the concept of implicit trust and adopts a philosophy of continuous verification, IoT devices often fail to reach the level of trust that ZT requires. 

Additional challenges arise with the second approach — integrating IoT into larger ZT-based systems. Most ZT security architectures rely on centralized knowledge management of the entities that compose a system, consistent with NIST recommendations \cite{NIST}. The decentralized and short-lived nature of IoT devices — combined with their heterogeneity and sheer number — adds complexity to their integration into ZT architectures. 
IoT integration into the ZT platform then necessitates tailored strategies and solutions. 

Obtaining a comprehensive overview of how this convergence is occurring in industry is complex. Mainstream ZT solutions are primarily offered by large vendors (such as GAFAM), since providing ZT capabilities requires developing and integrating numerous components into a consistent platform. These companies offer varying levels of transparency about their tools and differing support for ZT, IoT security, and their convergence \cite{DBLP:conf/cesar/Bobelin23}. 


ZT’s status as a widely used term complicates market analysis. The situation is further complicated because many so-called ZT or IoT solutions do not actually implement either concept in practice. Academic publications are likewise influenced by terminological trends; some authors frame their work as addressing ZT or IoT even when the connection is limited. The contribution of this paper is to give a better understanding of this domain, by answering the question:
\begin{quote}
   \textit{What is the current state of the art and evolution of IoT security and ZT convergence?}   
\end{quote}

To answer this question, this paper presents the first  Multivocal Literature Review (MLR)\cite{DBLP:journals/infsof/GarousiFM19} on this topic. An MLR is a type of Systematic Literature Review (SLR) \cite{SLR,KITCHENHAM20097} that encompasses both academic and industrial sources. An SLR applies a systematic methodology to review literature in a reproducible manner. MLR extends SLR by including the industrial literature. This MLR followed the widely adopted PRISMA 2020 framework guidelines for meta reviews \cite{Page2021PRISMA} for the current reporting of this work, while using Garousi and al \cite{DBLP:journals/infsof/GarousiFM19} process for conducting the MLR. 

Prior reviews on ZT such as \cite{Buck2021}, \cite{icaart24}, \cite{ITODO2024103827}, focus on specific subtopics (implementation, Intrusion Detection Systems, for example) related to IoT security and ZT convergence. None of them encompasses industrial and academic point of view on IoT security convergence with ZT to have an overview of the current state of the art. 

Key contributions of this paper are:
\begin{itemize}
    \item First MLR on IoT security–ZT convergence.
    \item Comparative analysis of academic vs industrial perspectives.
    \item Identification of research gaps. 
\end{itemize}

%
The remainder of the paper is structured as follows. Section \ref{S:context} explains the basics of ZT, IoT security, and the main ideas behind IoT security and ZT convergence. Section \ref{S:RelatedWork} gives an overview of work related to this paper. Then, Section \ref{S:methodology} describes this MLR approach to select and analyze literature, criteria for inclusion, search strategies, data extraction methods, and Research Questions (RQs) structuring the study. Section \ref{S:DataAnalysis} provides an analysis of the collected data, and Section \ref{S:results} answers the RQs. Section \ref{S:discussion} provides avenue for future research. Section \ref{S:threats} summarizes the threats to the validity of our study. Finally, Section \ref{S:conclusion} provides concluding remarks. 

\section{Background}
\label{S:context}
\subsection{ZT Overview}
ZT is a security model relying on the idea that perimeter-based security is inefficient when the so-called perimeter is breached;  nowadays, as phishing campaigns are more and more common, a user will likely compromise at least one of the resources enclosed within a perimeter. From an initial compromise of a single host, an attacker uses lateral movement—enabled by credential harvesting and remote execution—to progressively spread malware and gain control over a whole system by compromising adjacent machines, then networks, by compromising highly privileged accounts. ZT addresses the problem by requiring to never grant trust to other resources by default, thus forbidding lateral movements.

In practice, ZT is commonly understood either as:
\begin{enumerate}
    \item A motto/model/\textbf{\textit{set of guidelines}}, or
    \item \textbf{\textit{A type of security architecture}} formally defined by NIST \cite{NIST} and known as Zero Trust Architectures (ZTA).
\end{enumerate}

\subsubsection{\textbf{ZT Guidelines}}
\label{SS:ZT_pillar}
There is a consensus on the set of guidelines to follow \cite{ZTNA}:
\begin{itemize}
    \item All network flows MUST be authenticated before being processed.
    \item All network flows SHOULD be encrypted before being transmitted.
    \item Authentication and encryption MUST be performed by the endpoints in the network.
    \item All network flows MUST be enumerated so that access can be enforced by the system.
    \item The strongest authentication and encryption suites SHOULD be used within the network.
    \item Authentication SHOULD NOT rely on public PKI providers. Private PKI systems should be used instead.
    \item Devices SHOULD be regularly scanned, patched, and rotated.
\end{itemize}

\subsubsection{\textbf{ZT Architecture}}
\label{SS:NIST}
\begin{figure*}[!ht]
    \centering
    \includegraphics[width=\linewidth]{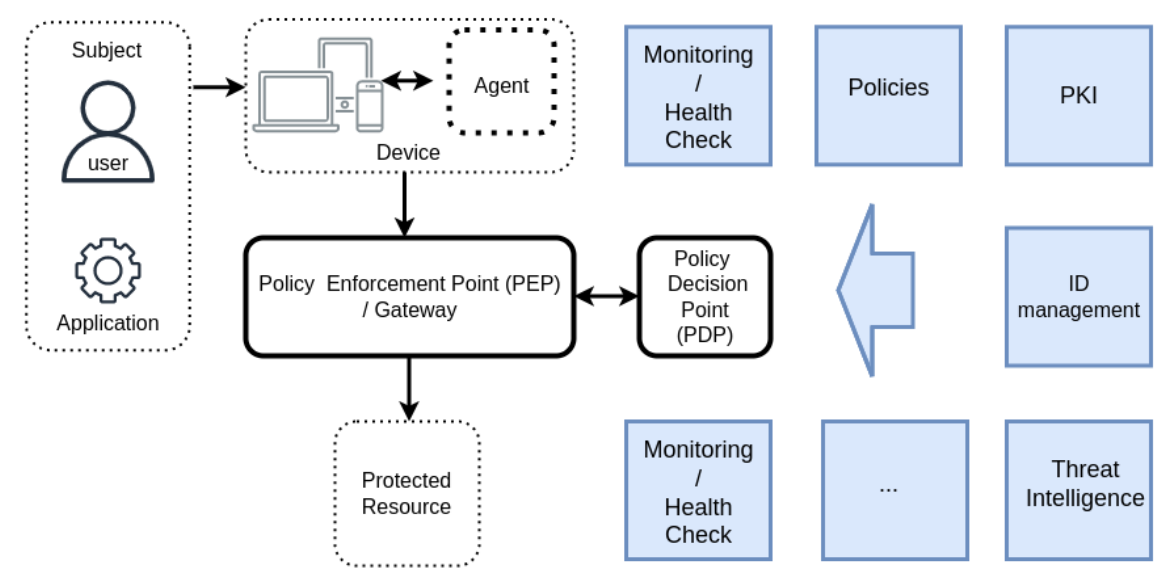}
    \caption{ZT Architecture (PIP in blue, core component with thick lines)}
    \label{Fig:ZTA}
    \Description{The subject, the resources, the data plane, the control plane, and the data sources used to make security decisions (PIP in blue, core component with thick lines).}
\end{figure*}

To comply with ZT guidelines, NIST recommended the logical architecture depicted inFigure \ref{Fig:ZTA}.

The standard identifies the issuer of the request as a subject that can be either a (human) user or an application/service. This subject uses a device to issue the request. This device may or may not be hosting a software agent, part of the ZT architecture, that will secure the asset and information provided by this device. The request for access targets a protected resource (that may be thought of as data or a service).
Policy Enforcement Point (PEP) is often implemented as a gateway: it enforces decisions about whether or not to grant trust to a flow by the Policy Decision Point (PDP). The separation between PEP and PDP relies on the separation between the control plane (which makes decisions on how to handle the traffic) and the data plane. PEP belongs to the data plane, while PDP relies on the control plane. The PDP decision-making is done using as many data sources as possible to make the wisest decision: information about the system state, the users, policies deployed, threat intelligence, etc. Those sources are named Policy Information Points (PIP). 

PDP itself in the NIST standard includes the Policy Engine (PE) component, which is the decision-making component, and the Policy Administration (PA) component, which is responsible for coordinating the actions of the PEP to reflect the decisions of the PE. Some authors and companies (see for instance [6]) add a Trust Engine component (TE). TE is responsible for running a Trust Algorithm (TA), interacting with the different data sources to evaluate risk. PE in this case makes its decision based on the risk evaluation returned by TE and the policies applying to the system. It is then not responsible for evaluating the risk per se. Google ZT solution BeyondCorp has pioneered the use of TE: it helps maintain a lower complexity of the system policy, by discarding edge cases and other unknown/unaddressed cases.

\subsection{IoT Security Overview}
\label{ss:IOTSec}

IoT devices' exposure stems from both intrinsic constraints—limited energy, computation, and storage—and extrinsic conditions such as physical accessibility and non-expert administration in consumer contexts (e.g., smart homes, smart farming, to name a few)~\cite{KOUICEM2018199,iotSoT,CPS_Survey,rfc8576}.


\begin{figure*}[htbp]
\centering
\begin{tikzpicture}[
    scale=0.05,
    layer/.style={
        rectangle, rounded corners, draw=black!70, thick, align=center,
        minimum width=9cm, minimum height=2cm, font=\small
    },
    zt/.style={
        rectangle, draw=black!80, thick, fill=blue!15,
        font=\footnotesize, align=center, minimum width=5cm, minimum height=2cm
    },
    every node/.style={inner sep=6pt},
    node distance=0.2cm
]

\definecolor{appblue}{RGB}{124,124,124}
\definecolor{midgreen}{RGB}{146,146,146}
\definecolor{netyellow}{RGB}{182,182,182}
\definecolor{sensorange}{RGB}{123,123,123}

\node[layer, fill=appblue] (app) {
    \textcolor{white}{\textbf{Application Layer}} \\[2pt]
    \textcolor{white}{\textit{Smart Services, Data Analytics, User Interfaces, Cloud Platforms}} \\[2pt]
    \textcolor{white}{Examples: Smart city dashboards, healthcare apps}
};

\node[layer, fill=midgreen, below=of app] (mid) {
    \textcolor{white}{\textbf{Middleware Layer}} \\[2pt]
    \textcolor{white}{\textit{Data Processing, Storage, API Management, Context Awareness}} \\[2pt]
    \textcolor{white}{Examples: IoT brokers (MQTT, CoAP), edge nodes, message queues}
};

\node[layer, fill=netyellow, below=of mid] (net) {
    \textcolor{white}{\textbf{Network Layer}} \\[2pt]
    \textcolor{white}{\textit{Communication Infrastructure, Protocols, Gateways, Security Controls}} \\[2pt]
    \textcolor{white}{Examples: 5G, Wi-Fi, ZigBee, IPv6, VPNs, routing, firewalls}
};

\node[layer, fill=sensorange, below=of net] (sens) {
    \textcolor{white}{\textbf{Perception (Sensing) Layer}} \\[2pt]
    \textcolor{white}{\textit{Sensors, Actuators, RFID, Embedded Controllers}} \\[2pt]
    \textcolor{white}{Examples: Temperature sensors, RFID tags, motion detectors}
};


\node[zt, right=1.5cm of app.east] (pdp) {\textbf{Policy Decision Point (PDP)} \\ Access Policies, Risk Engine};
\node[zt, right=1.5cm of mid.east] (pep) {\textbf{Policy Enforcement Point (PEP)} \\ API Gateway, Broker Security};
\node[zt, right=1.5cm of net.east] (idp) {\textbf{ID Management} \\ IAM, MFA, Certificates};
\node[zt, right=1.5cm of sens.east] (telemetry) {\textbf{Policy Information Point (PIP)}\\ Telemetry, Continuous Monitoring};

\draw[<->, thick, black!60, dashed] (app.east) -- (pdp.west);
\draw[<->, thick, black!60, dashed] (mid.east) -- (pep.west);
\draw[<->, thick, black!60, dashed] (net.east) -- (idp.west);
\draw[<->, thick, black!60, dashed] (sens.east) -- (telemetry.west);



\end{tikzpicture}
\caption{Four-layer IoT architecture (on the left) with mapped Zero Trust (ZT) control components (on the right). Each layer interacts with corresponding ZT elements to enforce continuous authentication, policy decision-making, and telemetry-driven trust evaluation.}
\label{fig:iot-zt-architecture}
\Description{Four-layer IoT architecture (on the left) with mapped Zero Trust (ZT) control components (on the right). Each layer interacts with corresponding ZT elements to enforce continuous authentication, policy decision-making, and telemetry-driven trust evaluation.}
\end{figure*}
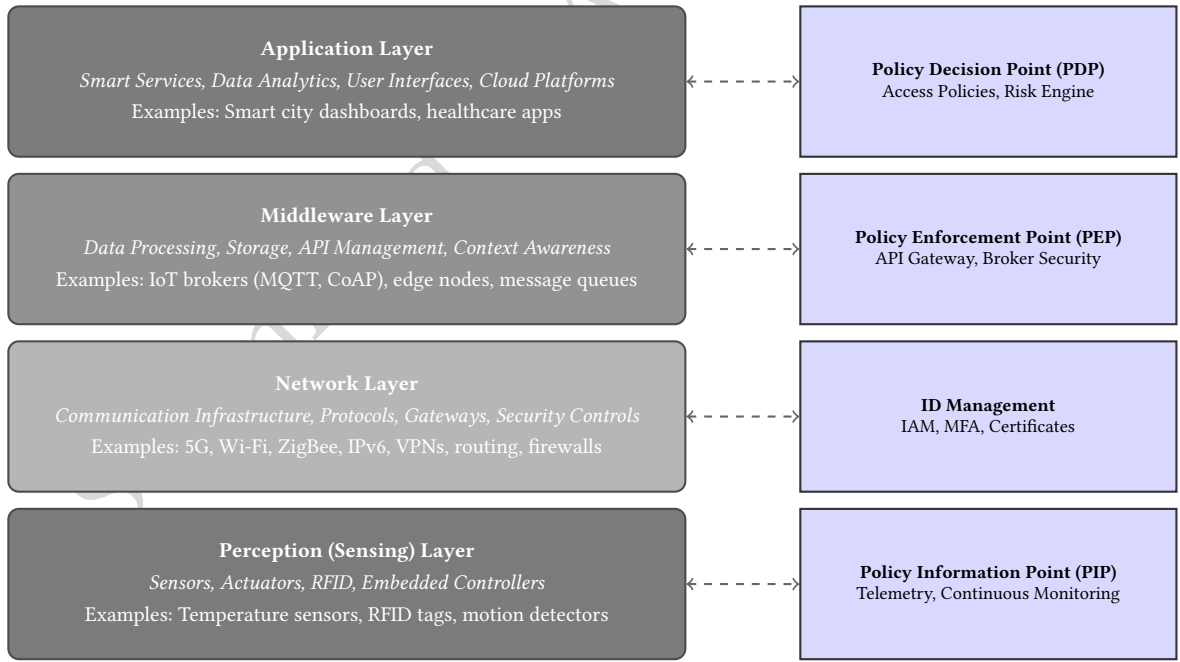

While IoT security borrows concepts from general device protection—encryption, secure boot, patch management, and access control—it remains uniquely constrained by heterogeneity, decentralization, and limited hardware capabilities. Consequently, IoT security enforcement adopts tailored protocols and lightweight mechanisms rather than replicating enterprise-level models. To do so, IoT security mechanisms spans the four  typical IoT architecture layers illustrated by boxes on the left in Figure \ref{fig:iot-zt-architecture}: (1) \emph{perception} (sensing and actuation), (2) \emph{network} (communication), (3) \emph{middleware} (data aggregation and storage), and (4) \emph{application} (analytics and services). Each layer introduces distinct attack vectors and protection requirements. 

\subsection*{IoT Security Integration}
\label{ss:threatsonsystems}
When IoT components integrate with broader infrastructures, they expand the attack surface to both endpoints and management platforms. Common threats include\cite{IoTCloudIntegration2023}:
\begin{itemize}
  \item Denial of service via device unavailability,
  \item Data poisoning or sensor cloning,
  \item Information leakage through unsecured communications, and
  \item System compromise by exploiting IoT as an initial intrusion vector.
\end{itemize}

Mitigation relies on comprehensive security ecosystems offering capabilities such as:
\begin{itemize}
  \item {Asset discovery and vulnerability scanning} to identify connected devices and weak configurations \cite{miettinen2017iot-sentinel}, \cite{yu2020survey-vuln-iot}, \cite{ferrara2021static-analysis-iot},
  \item {Device authentication and authorization} to control access based on identity and privileges,
  \item {Secure communication} through encryption protocols (e.g., TLS/SSL),
  \item {Continuous monitoring and threat detection} for anomaly identification, and
  \item {Patch management and vulnerability assessment} for timely remediation~\cite{rfc8576}.
\end{itemize}

Ultimately, securing IoT ecosystems requires harmonizing these operational controls with ZT principles—ensuring continuous verification and context-aware enforcement across highly heterogeneous environments.

Figure \ref{fig:iot-zt-architecture} illustrates how ZT elements defined in the NIST architecture (on the right) interact with the four foundational layers of the IoT architecture (on the left). At the application level, policy decision points (PDPs) evaluate contextual access requests; in the middleware, policy enforcement points (PEPs) implement those decisions and regulate data exchange through APIs and brokers. The network layer relies on identity providers (IdPs) to authenticate entities and manage cryptographic credentials, while the perception layer integrates telemetry and device-trust mechanisms that continuously assess sensor integrity, and thus provide information, and then may be considered as a PIP. 
Effective convergence requires synchronized advances in technical infrastructure, cryptographic agility, and workforce competencies, and this effort can be done in two ways: adapt ZT architecture to IoT, or adapt IoT to ZT architecture. 

\subsection{IoT Security and ZT Convergence}
\begin{figure*} [t]
    \centering
    \includegraphics[width=1\linewidth]{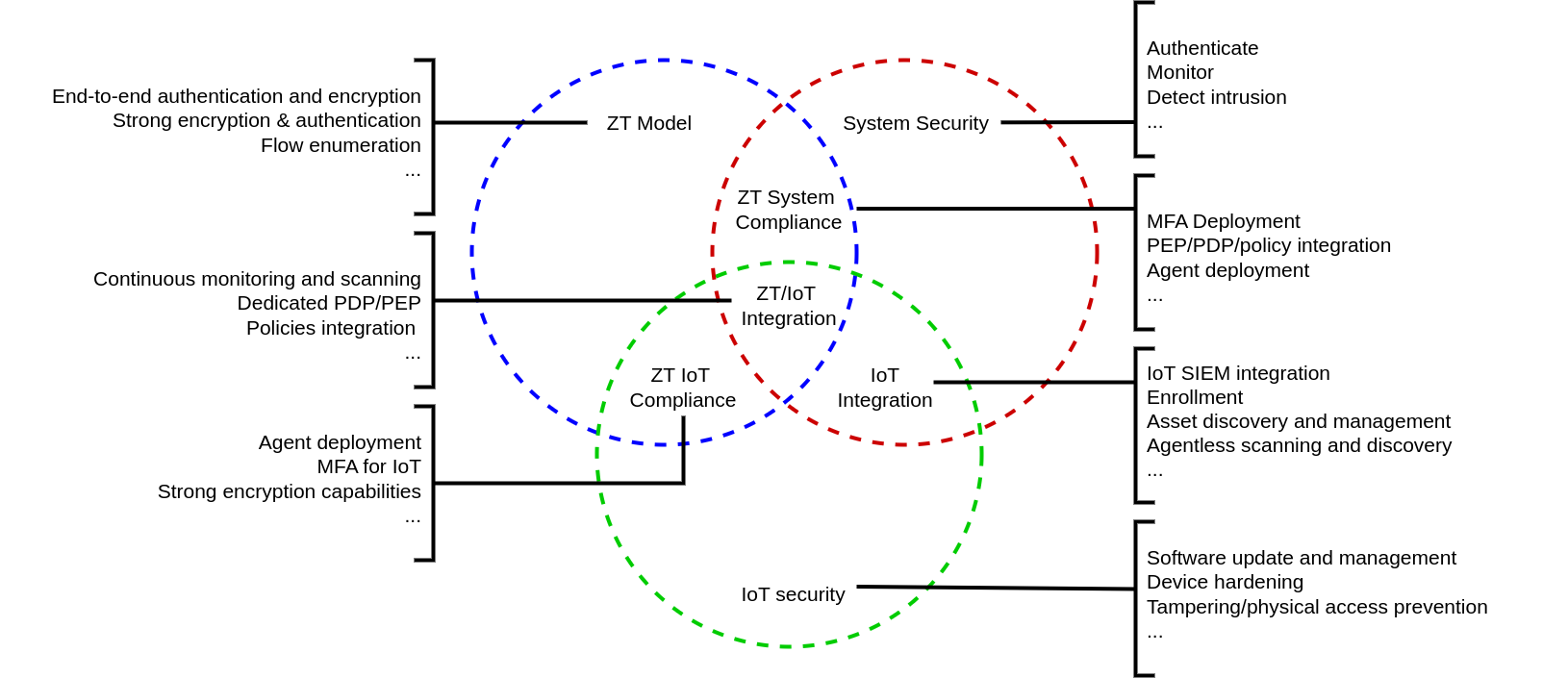}
    \caption{Integration of IoT security into ZT}
    \label{Fig:IoTintegration}
    \Description{The conjunction of 3 different fields: system security, ZT security model and IoT security.}
\end{figure*}

An overview of the problem of securing systems containing IoT using ZT security model is given in Figure \ref{Fig:IoTintegration}. It can be seen as the conjunction of 3 different fields, and decomposed into different topics: making the system security compliant with ZT, making IoT compliant with ZT, and integrating IoT (sub)system into the whole system security. 
This paper deals only with topics related to ZT and IoT security convergence. Securing IoT devices themselves, securing the system itself, refining/defining the ZT model, and system compliance with ZT requirements, are all out of the scope of this study. 

However, two different topics are crucial: (1) IoT compliance with ZT requirements, (2) IoT and ZT integration. The former deals with how the IoT device may be refactored, modified, or configured to be able to be included in the ZT architecture, or at least to embrace its philosophy. An example of work in this paper includes the use of blockchain to increase IoT trust \cite{Li2023}. The latter deals with how the ZT model may be changed to take into account the IoT characteristics, or how specific tools or mechanisms may be put in place to integrate IoT into a ZT platform. Isolation for example is the solution advocated by most of the industry falling into that topic \cite{DBLP:conf/cesar/Bobelin23}. This paper deals with academics and industrial solutions for both of those interrelated topics. 

\section{Related Work}
\label{S:RelatedWork}
As stated before, there is no -up to our knowledge- MLR related to the subject of IoT security and ZT convergence. 

IoT security has been an active subject for many years, and therefore many SLRs have been published either directly on IoT security \cite{Bekkali}, \cite{Rajmohan2022}, or more specifically on IoT forensic \cite{AKINBI2022301470} for example. \cite{electronics13244965} deals with generative IA applications to IoT security. Other SLRs that deal with IoT may contain sections dedicated to security, for example, \cite{9825678} deals with IoT in general, \cite{BENIWAL20229541} focuses on IoT gateways, \cite{9961212} focuses on lightweight blockchains for IoT, \cite{electronics11193223} focuses on IoT applications in healthcare. \cite{Colombo2021} is a state-of-the-art on access control enforcement in IoT. 

Much work exists for ZT: \cite{icaart24}, for example, is an SLR focusing on Intrusion Detection Systems, and ZT, \cite{ITODO2024103827} is an MLR focusing on implementation aspects of ZT. Some of them discuss, to a certain extent, IoT security and ZT convergence. 
 Another state-of-the-art focuses on ZT \cite{Liu2024} and then considers how ZT may be implemented given the IoT context.  An SLR recently published focuses on ZT and its possible use in IoT, but does not encompass the industrial work \cite{AZAD2024101227}. \cite{Annabi_2024} surveys ZT in the context of autonomous vehicles. Industrial solutions that integrate IoT in ZT are reviewed in \cite{DBLP:conf/cesar/Bobelin23}. This work targets only complete solutions provided by major industrial actors, with a specific focus on estimating how strong the support for IoT integration into available products is. \cite{Buck2021} is an MLR focusing on ZT that includes some statements on IoT; \cite{Syed2022} is a survey on ZT with some discussion about the challenges IoT induces, as well as \cite{Yan2020}, \cite{e25121595}, and \cite{ghasemshirazi2023zerotrustapplicationschallenges}. \cite{Tanque2023} is a book chapter discussing ZT and IoT. \cite{JAIN2023} is a handbook on IoT security that integrates a small chapter on IoT integration into ZT. \cite{Li2019} is an editorial that introduces ZT for IoT. \cite{Li2022} introduces challenges induced by ZT in 5G/6G environments. \cite{10680244} focuses on ZT implementations for IIoT. 
 
 Table \ref{latable} gives a summary of the related work and how they relate to IoT, ZT, academic work, industrial aspects, and if they used a Systematization of Knowledge methodology (SLR, MLR or other systematic approaches). 

  \begin{table*}[t!]
\centering
\footnotesize   
\caption{Condensed Classification of Related Work References by Topic (IoT, Zero Trust, IoT + ZT)}
\label{latable}
\begin{tabular}{|l|p{6cm}|c|c|c|c|c|}
\hline
\textbf{Citation} & \textbf{Title} & \textbf{IoT} & \textbf{ZT} & \textbf{Acad.} & \textbf{Ind.} & \textbf{Syst.} \\
\hline
\rowcolor[gray]{0.9}
\multicolumn{7}{|c|}{\textbf{IoT-focused Systematic Reviews}} \\
\hline
\cite{Bekkali} & Systematic Literature Review of IoT Security & \checkmark &  & \checkmark &  & \checkmark \\
\cite{Rajmohan2022} & Patterns and Architectures for IoT Security & \checkmark &  & \checkmark &  & \checkmark \\
\cite{AKINBI2022301470} & Blockchain-based IoT Forensic Investigation Models & \checkmark &  & \checkmark &  & \checkmark \\
\cite{electronics13244965} & Generative AI Solutions in IoT Security & \checkmark &  & \checkmark &  & \checkmark \\
\cite{9825678} & The Anatomy of IoT Platforms — Multivocal Mapping & \checkmark &  & \checkmark &  & \checkmark \\
\cite{BENIWAL20229541} & Systematic Review on IoT Gateways & \checkmark &  & \checkmark &  & \checkmark \\
\cite{9961212} & Lightweight Blockchain for IoT: SLR & \checkmark &  & \checkmark &  & \checkmark \\
\cite{electronics11193223} & Enabling IoT in Healthcare: Motivations, Challenges, and Recommendations & \checkmark &  & \checkmark &  & \checkmark \\
\hline
\rowcolor[gray]{0.9}
\multicolumn{7}{|c|}{\textbf{Zero Trust (ZT)-focused Studies}} \\
\hline
\cite{icaart24} & Zero Trust for Intrusion Detection Systems: SLR &  & \checkmark & \checkmark &  & \checkmark \\
\cite{ITODO2024103827} & Multivocal Review on Zero-Trust Implementation &  & \checkmark & \checkmark &  & \checkmark \\
\cite{Buck2021} & Never Trust, Always Verify: A Multivocal Review of Zero-Trust &  & \checkmark & \checkmark &  & \checkmark \\
\cite{Syed2022} & Zero Trust Architecture (ZTA): Comprehensive Survey &  & \checkmark & \checkmark &  & \checkmark \\
\cite{ghasemshirazi2023zerotrustapplicationschallenges} & Zero Trust: Applications, Challenges, and Opportunities &  & \checkmark & \checkmark &  &  \\
\hline
\rowcolor[gray]{0.9}
\multicolumn{7}{|c|}{\textbf{IoT + ZT Convergence Studies}} \\
\hline
\cite{Colombo2021} & Access Control in IoT: Challenges in the Zero Trust Era & \checkmark & \checkmark & \checkmark &  &  \\
\cite{Liu2024} & Dissecting Zero Trust and Its Implementation in IoT & \checkmark & \checkmark & \checkmark &  & \checkmark \\
\cite{AZAD2024101227} & Verify and Trust: Zero-Trust Security in the Age of IoT & \checkmark & \checkmark & \checkmark &  & \checkmark \\
\cite{Annabi_2024} & Towards Zero Trust Security in Connected Vehicles & \checkmark & \checkmark & \checkmark &  & \checkmark \\
\cite{DBLP:conf/cesar/Bobelin23} & Industrial Literature Review: Zero Trust in IoT & \checkmark & \checkmark &  & \checkmark & \checkmark \\
\cite{Yan2020} & Survey on Zero-Trust Network Security & \checkmark & \checkmark & \checkmark &  &  \\
\cite{e25121595} & Theory and Application of Zero Trust Security & \checkmark & \checkmark & \checkmark &  & \checkmark \\
\cite{Tanque2023} & Cyber Risks on IoT Platforms and Zero Trust Solutions & \checkmark & \checkmark & \checkmark &  &  \\
\cite{JAIN2023} & IoT and OT Security Handbook & \checkmark & \checkmark &  & \checkmark &  \\
\cite{Li2019} & Editorial: Zero Trust based Internet of Things & \checkmark & \checkmark & \checkmark &  &  \\
\cite{Li2022} & Future Industry IoT with Zero-trust Security & \checkmark & \checkmark & \checkmark &  &  \\
\cite{10680244} & Shared Responsibility and Zero Trust in the Industrial IoT & \checkmark & \checkmark & \checkmark &  &  \\
\hline
\rowcolor[gray]{0.7}
This work & Converging Zero Trust and IoT Security: A MLR & \checkmark & \checkmark & \checkmark & \checkmark & \checkmark \\
\hline
\end{tabular}
\end{table*}



\section{Methodology}
\label{S:methodology}
The industrial sector plays a leading role in research on the convergence of ZT and IoT security. To analyze existing knowledge from both academic and industrial perspectives, a Multivocal Literature Review (MLR) \cite{DBLP:journals/infsof/GarousiFM19} was conducted. The review followed Garousi et al.’s methodology \cite{GarousiFM17} and PRISMA 2020 guidelines \cite{Page2021PRISMA}. More details about our methodology is provided in appendix \ref{S:A:methodology}.

MLR combines Academic Literature (AL), which includes peer-reviewed scholarly works, with Grey Literature (GL), such as technical reports, white papers, and blogs. While GL is less formal, it often provides practical insights and recent developments. In a MLR, sources are categorized in tiers by their credibility and level of outlet control: Tier 1 (High) for white papers, tier 2 (Medium) for news articles and corporate reports, and Tier 3 (Low) for social media. To ensure relevance, only high and medium quality GL (tiers 1 and 2) were considered in this study. Tier 1 includes sources like NIST SP 800-207 \cite{NIST} and industry white papers such as Microsoft Azure's white paper \textit{Zero Trust Cybersecurity for the Internet of Things} \cite{MicrosoftWhitePaper}. 

\subsection{Research Questions and Data Analysis Strategy}
\begin{table*}[h]
\caption{Research Questions Summarized}
\label{tableQ}
\begin{tabular}{ p{2cm} p{3.5cm} p{3.5cm} p{3.5cm}  }
     \toprule
 Level of analysis & Design and features & Measurement and value & Management and organization \\  
 \midrule

 Concept and architecture &  RQ1: How to technically realize convergence of IoT security and ZT? & RQ2: What are the benefits and limitations of IoT security and ZT convergence? &  RQ3: What are the human resources needed to maintain a system that adopted a ZT and IoT security convergence philosophy? \\  

 Firms and industries & RQ4: How can organizations realize IoT security and ZT convergence? & RQ5: How do IoT security and ZT convergence provide added value for organizations? & RQ6: How should organizations organize, govern, fund, and develop IoT security and ZT convergence? \\  

 Users and society & RQ7: How do IoT security and ZT convergence affect the interaction between users and technology? &  RQ8: What are the benefits and costs of ZT and IoT security convergence for individual users and society? & RQ9: How does one balance user privacy and ZT and IoT security convergence requirements? \\  
 \bottomrule
\end{tabular}

\end{table*}

This review used and adapted the research questions framework already appearing in former MLRs such as \cite{Buck2021}, \cite{risius}, or \cite{61d3893a-0516-3103-b07e-cb4fc06d28ac}. The basic idea of this framework is to consider the literature from two different perspectives they named \textit{dimensions}.

Those two dimensions are \textit{activities} and \textit{levels of analysis} influenced by activities. Activities pertain to the actions that developers or users undertake and are categorized into three groups. First, \textit{design and feature} activities involve understanding the implementation and design of concepts, including the consequences of certain design choices. Second, \textit{measurement and value} questions revolve around the added value provided, specifically how to create and measure additional value for stakeholders. Third, \textit{management and organization} encompass actions necessary for successful implementation, including required organizational capabilities, skills, talents, and management of sensitive governance-related aspects. 

The \textit{level of analysis} specifies the scope of the research object. This review used the level of analysis defined in \cite{Buck2021}. 
This review consider the \textit{concept and architecture} level, focusing on architectural variations and protocols, \textit{Firms and industries} level focusing on organizations, and \textit{users and society} level focusing on end users and societal issues. 
For each couple of levels and activities, questions to be answered were defined; those are listed in Table \ref{tableQ}.

\subsection{Search Strategies}

GL and AL artifacts to analyze were searched concurrently on academic databases (IEEE Xplore, ACM Digital Library, Science Direct, and Google Scholar) and using Google search engine for GL. The searches were looking for documents containing both  keywords relevant for IoT and ZT ("IoT" OR "Internet Of Things", "ZT" OR "Zero Trust"), published since 2014. The searches were performed during a period between November 2023 and March 2024. The full process is described in appendix \ref{S:A:searches}. Figure \ref{Fig:searchstrategy} provides an outline of this study search strategies and results, as a PRISMA flow diagram \cite{Page2021PRISMA}. 

\begin{figure*} [!ht]
    \begin{center}
    \includegraphics[width=\linewidth]{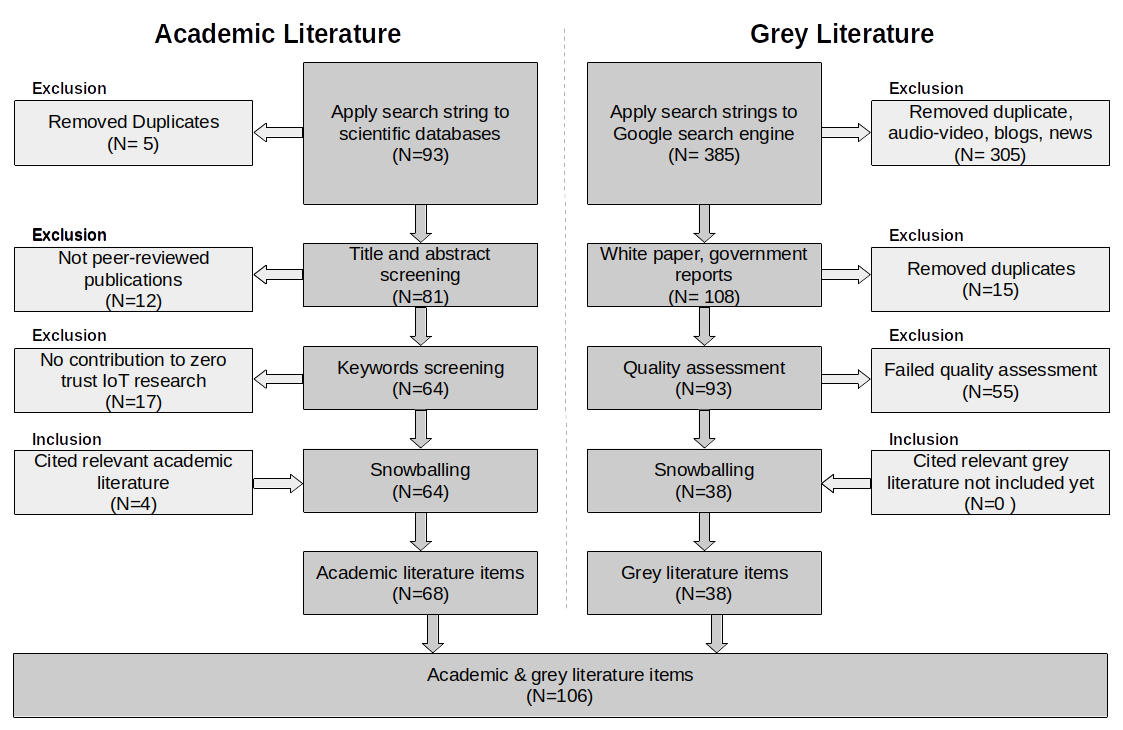}
    \end{center}
\caption{PRISMA flow diagram (AL on the left, GL on the right)}
\Description{The number of academic and industrial items (included and excluded) used for this MLR study (AL on the left, GL on the right)}
\label{Fig:searchstrategy}
\end{figure*}

\section{Preliminary Data Analysis}
\label{S:DataAnalysis}

\begin{figure*} [ht]
    \begin{center}
    \includegraphics[width=0.49\linewidth]{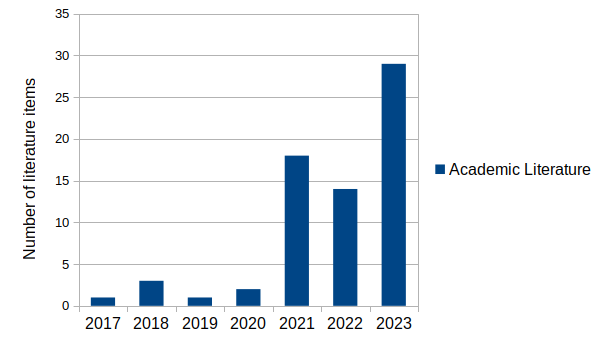}
    \includegraphics[width=0.49\linewidth]{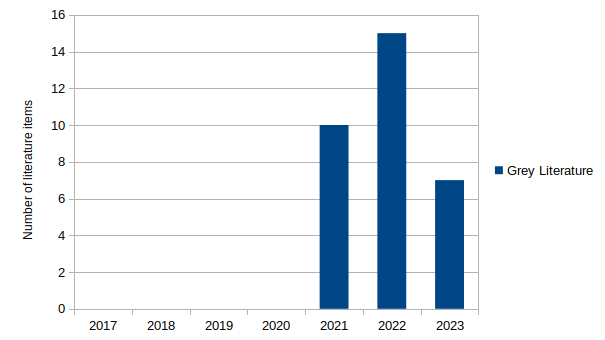}
    \end{center}
\caption{Distribution of AL and GL over time (AL on the left, GL on the right side)}
\Description{The number of searched academic and grey literature items per year (AL on the left, GL on the right side)}
\label{Fig:academicliterature}
\end{figure*}

Figure \ref{Fig:academicliterature} illustrates the distribution of AL and GL items over time and shows that there are just a few academic papers that address IoT security and ZT convergence before 2021. The first academic paper dealing with this topic was published in 2017. This is consistent with the timeline of the growing interest in ZT observed in previous SLRs such as \cite{Buck2021}.

\begin{figure*} [!ht]
    \begin{center}
    \includegraphics[width=\linewidth]{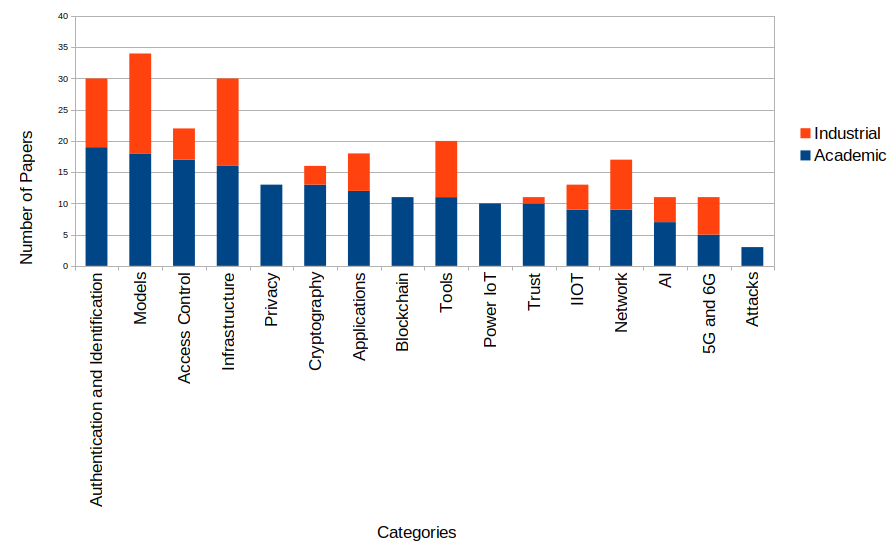}
    \end{center}
\caption{Distribution of papers per topic (AL in blue, and GL in red)}
\Description{The number of papers that deal with the categorized topics for both AL (in blue) and GL (in red).}
\label{Fig:art_cat}
\end{figure*}

An analysis based on the topics of both AL and GL was also done. To do so, after reviewing papers, an identification of topics shared by multiple papers was performed. Details about the topic description and how it is considered in both AL and GL is given in Appendix \ref{S:AppendixDataAnalysis} 
Figure \ref{Fig:art_cat} shows the number of papers that deal with those topics, for both AL and GL. The first observation is that there is no strong predominance of a particular subject. This topic diversity is expected given the broad scope of ZT and IoT security convergence, and is consistent with previous studies that dealt only with AL, such as \cite{AZAD2024101227}, confirming the accuracy of the dataset the search methodology produced. This Figure also shows a preliminary result: as the study also addresses GL contrary to the previous one, it show that some topics are addressed only by the AL and not by the industry (privacy, power IoT and attacks). 

The Tables \ref{tableRQAL} and \ref{tableRQGL} give the number of relevant papers for each RQs for AL and GL respectively. One immediate conclusion is that most of the work, in both GL and AL, is focused on \textit{design and features} activities at the \textit{concept and architecture} level of analysis. On the other hand, the user and society level, and the management and organization activities remain almost unexplored research fields.

\begin{table*}[t]



\caption{AL mapping to RQ}
\label{tableRQAL}
\begin{tabular}{l|c|c|c}
    \toprule
    \textbf{Level of analysis} & \textbf{Design and features} & \textbf{Measurement and value} & \textbf{Management and organization} \\
    \midrule
    Concept and architecture & \cellcolor{gray!100} 60 & \cellcolor{gray!10} 3 & \cellcolor{gray!0} 0 \\
    \hline
    Firms and industries & \cellcolor{gray!55} 17 & \cellcolor{gray!3} 1 & \cellcolor{gray!6} 2 \\
    \hline
    Users and society & \cellcolor{gray!0} 0 & \cellcolor{gray!6} 2 & \cellcolor{gray!3} 1 \\
    \bottomrule
\end{tabular}

\quad



\caption{GL mapping to RQ}
\label{tableRQGL}
\begin{tabular}{l|c|c|c}
    \toprule
    \textbf{Level of analysis} & \textbf{Design and features} & \textbf{Measurement and value} & \textbf{Management and organization} \\
    \midrule
    Concept and architecture & \cellcolor{gray!100} 37 & \cellcolor{gray!11} 4 & \cellcolor{gray!0} 0 \\
    \hline
    Firms and industries & \cellcolor{gray!41} 15 & \cellcolor{gray!92} 34 & \cellcolor{gray!0} 0 \\
    \hline
    Users and society & \cellcolor{gray!0} 0 & \cellcolor{gray!0} 0 & \cellcolor{gray!0} 0 \\
    \bottomrule
\end{tabular}

\end{table*}

\section{Results}
\label{S:results}
This section gives answers to the different RQ defined in Section \ref{S:methodology}. The answers rely on an in-depth study of the corpus artifacts given in Appendix \ref{S:AppendixMapping}, by providing a detailed analysis of most relevant artifacts for each RQ. 

\subsection{Level of Analysis: Concept and Architecture}
\subsubsection{RQ1: Technical convergence of IoT security and ZT }
\label{RQ1}
As discussed in Section~\ref{S:context}, two distinct approaches to the convergence of IoT security and Zero Trust (ZT) can be identified: (1) adapting IoT systems to comply with ZT principles, and (2) integrating IoT components within an organization-wide ZT architecture.

Regarding Academic Literature (AL), most studies emphasize the first approach—aligning IoT design and operation with ZT philosophy. These works focus on technical adaptations such as lightweight encryption, decentralized identity management, and context-aware trust mechanisms that enable IoT devices to function as native actors within a ZT ecosystem. In contrast, Grey Literature (GL) predominantly supports the second approach, prioritizing the pragmatic integration of IoT into existing enterprise ZT frameworks. Industrial reports and white papers tend to assume that IoT systems already exist in production and therefore explore methods to extend ZT controls—such as identity, credential, and access management (ICAM) or policy enforcement points (PEPs)—to heterogeneous and resource-constrained IoT environments.

This divergence constitutes a key finding of the present study: \textbf{academic and industrial sources fundamentally differ in their vision of how IoT security should converge with ZT principles}. Academic works largely propose reengineering IoT devices and protocols to make them inherently ZT-compliant, while industrial sources focus on incremental integration into enterprise-grade ZT implementations guided by NIST SP~800-207.

The emphasis observed in GL reflects the operational reality of complex, brownfield IoT deployments, where full redesign is rarely feasible. These works operate under realistic assumptions about device heterogeneity, legacy constraints, and limited computational capabilities. Consequently, they advocate a gradual convergence path—embedding IoT within NIST-compliant ZT architectures and leveraging centralized policy decision mechanisms. This practical orientation also reveals the slow adoption of emerging security standards and limited organizational agility in evolving security postures~\cite{NISTCryptoAgility,rosserlarri}.

Conversely, academic perspectives remain largely speculative, addressing next-generation IoT paradigms that could natively embody ZT principles through architectural redesigns and cryptographic innovation. While this line of inquiry advances theoretical understanding, it is often detached from immediate industrial applicability. Bridging this gap between conceptual design and operational integration remains an open challenge and a promising direction for future research.

\subsubsection{RQ2: Technical benefits and limitations of IoT security and ZT convergence}

Only limited work in either the academic literature (AL) or grey literature (GL) explicitly quantifies the benefits and limitations of IoT security and Zero Trust (ZT) convergence. Assessing the trade-offs and risks associated with the two dominant convergence strategies—(1) making IoT systems compliant with ZT principles, and (2) integrating IoT into existing ZT architectures—remains an open research area. Yet this question is central to the adoption of ZT, as perceived benefits, implementation costs, and organizational acceptability strongly influence decision-making~\cite{Buck2021}.

While several studies examine ZT adoption in general organizational contexts~\cite{Liu2024,YEOH2023103412,ghasemshirazi2023zerotrustapplicationschallenges,Lund2024ZeroTrust}, none isolate or quantify the specific cost–benefit dynamics of ZT–IoT convergence. Existing evaluations typically focus on overall ZT return-on-investment, performance, or risk reduction, but not on how IoT integration modifies those metrics. Consequently, \textbf{no empirical or comparative study has yet evaluated the concrete advantages, trade-offs, or unintended consequences of converging IoT security and ZT principles}. Establishing such evidence is essential for guiding both academic design choices and industrial deployment strategies.

\subsubsection{RQ3: Human resources needed to maintain a system that adopted a ZT and IoT security convergence philosophy}

Neither AL nor GL provides a detailed analysis of workforce requirements for sustaining ZT–IoT convergence. Although some industrial materials include training components, no formal studies assess the competencies or organizational capacities necessary to operate converged systems. Transitioning from perimeter-based security to ZT already demands expertise spanning identity, network, and policy management~\cite{ZTNA}. Integrating IoT further amplifies this challenge by introducing device heterogeneity, embedded constraints, and operational technology (OT) considerations. \textbf{This lack of multidisciplinary expertise is a recognized factor in the slow adoption of ZT frameworks}~\cite{Liu2024,YEOH2023103412,ghasemshirazi2023zerotrustapplicationschallenges,Lund2024ZeroTrust}, and the complexity increases when IoT ecosystems are involved.

The NIST National Initiative for Cybersecurity Education (NICE) Workforce Framework~\cite{nistNICE2020} offers a useful taxonomy of knowledge, skills, and abilities (KSAs) for cybersecurity roles. Several NICE roles—such as those related to identity and access management, cryptographic key operations, and continuous monitoring—align directly with ZT functions. However, ZT–IoT convergence introduces interdisciplinary requirements that are underrepresented in current workforce models. Effective deployment and maintenance of distributed trust architectures call for professionals who can bridge OT, embedded systems, and enterprise IT domains.

This competency gap also extends to cryptographic agility, emphasized in the NIST \textit{Considerations for Achieving Cryptographic Agility} white paper~\cite{NISTCryptoAgility}. The ability to rapidly update cryptographic algorithms, keys, and protocols in response to evolving threats is becoming an operational necessity for ZT environments that integrate long-lived, resource-constrained IoT devices. It is also mandatory for an engineer managing the platform to understand interoperability problems that occurs when dealing with heterogeneous cryptographic protocols and implementations. Table~\ref{tab:zt_iot_workforce} summarizes the alignment between technical functions, workforce competencies, and relevant NIST guidance, highlighting the \textbf{need for new interdisciplinary training and certification pathways, to train expert able to manage the IoT security and ZT operational convergence.}


\begin{table*}[htbp]
\centering
\caption{Alignment between Technical Requirements, Workforce Competencies, and NIST Framework References for ZT--IoT Convergence}
\label{tab:zt_iot_workforce}
\begin{tabular}{@{}p{0.22\linewidth} p{0.33\linewidth} p{0.35\linewidth}@{}}
\toprule
\textbf{Technical Requirement (ZT--IoT)} & \textbf{Corresponding Workforce Competencies (NICE Roles / KSAs)} & \textbf{Relevant NIST Framework or Guidance} \\ 
\midrule
Identity, credential, and access management (ICAM) &
Identity and Access Management Specialist; Knowledge of authentication, authorization, and credential lifecycle management &
\textit{NICE SP~800--181~Rev.~1, ``Protect and Defend'' category; NIST SP~800--207 (Zero Trust Architecture)} \\[0.6em]

Cryptographic agility and key management &
Cryptographic Technician / Security Architect; Skills in algorithm migration, certificate management, and key rotation &
\textit{NIST CSWP~39, ``Considerations for Achieving Cryptographic Agility'' (2025)} \\[0.6em]

Continuous monitoring and trust evaluation &
Cyber Defense Analyst; Ability to analyze security telemetry and apply risk-based access control policies &
\textit{NICE ``Analyze'' and ``Operate \& Maintain'' categories; NIST SP~800--137 (Information Security Continuous Monitoring)} \\[0.6em]

IoT device posture assessment and attestation &
Systems Security Engineer; Understanding of embedded/OT systems and secure provisioning &
\textit{NICE ``Securely Provision'' category; NIST SP~800--213 (IoT Device Cybersecurity Guidance)} \\[0.6em]

Privacy and data minimization in telemetry &
Privacy Engineer; Competence in privacy-by-design and data governance controls &
\textit{NIST Privacy Framework (2020)} \\ 
\bottomrule
\end{tabular}
\end{table*}
 
 \subsection{Level of analysis: Firms and Industries}
 \subsubsection{RQ4: How can an organization realize IoT security and ZT convergence} 

Both AL and GL address this topic, but with different points of view. AL focuses mainly on specific industrial sector or application (healthcare, IIOT) and the impact of their specificity on the implementation of IoT security and ZT convergence, while GL is more oriented toward generic off-the-shelf solutions.\textbf{ AL analyzes the cost of adoption of converged solutions by sectors, while those costs may be underestimated by the industry, or hidden from their customers. }
 
\subsubsection{RQ5: How does IoT security and ZT convergence provide added value for organizations}
 
 In AL, there is little to no attempt to measure the added value for organizations about this convergence. In the corpus, only \cite{marketing} compares perimeter-based security and ZT in the context of IIoT. 

 In GL, this topic is massively discussed, as it is a key factor in the adoption of this convergence in the organization. Indeed, the promise of stronger security is in most of the GL reviewed. However, no artifacts address the question of concretely quantifying the gain associated with the IoT and ZT convergence. Demonstrating the added value of ZT and IoT security convergence is done by comparing it to segmentation-based security. 
 The drawbacks of such a convergence, often listed when dealing with ZT adoption (see for example \cite{ZTNA}), are not addressed in GL. 
 
 The organization-level choice of either strengthening IoT subsystem by making it comply with ZT philosophy or integrating IoT subsystem into ZT organization-wide system is never discussed in any of the literature. 

 In sum, \textbf{there is a lack of formal (independent) study about the costs and added value of this convergence in AL}, which would be undoubtedly of interest to organizations. 
 
 \subsubsection{RQ6: Organizations setup, government and development of IoT security and ZT convergence}
One may foresee the complexity of the process induced by a shift from an existing IoT security to the convergence of IoT and ZT by looking at the complexity of shifting from perimeter-based security to ZT. Shifting is a long process that usually lasts for years, carefully migrating segments one after the other, with meticulous transcription of policies \cite{ZTNA}. 

Neither AL nor GL dataset contains paper dealing with the subject of how an organization may manage these projects. Two papers in this study discuss factors influencing the organization of this process, focusing on healthcare systems. \cite{Samah2023} studies factors behind the adoption of ZT and IoT in the Malaysian healthcare system. \cite{Gofwen2023} analyses the impact on privacy of IoT integration in ZT, and the impact on adoption of IoT in healthcare systems. While those papers give insights into this domain, there is \textbf{no detailed definition of the process and verification of security assessment during transition to a ZT-based IoT security}.
 
 \subsection{Level of Analysis: Users and Society}
This study shares the same observation as a previous MLR focusing only on ZT \cite{Buck2021}: there are very few papers, either in AL or GL, that consider ZT and IoT from the users and society level of analysis. It could be either an artifact of the methodology of this study\footnote{Because of the literature considered, as the search methods used are oriented toward research (so, technical) papers. This is discussed in section \ref{S:threats}}, or a real lack of support for those aspects.  
 
\subsubsection{RQ7: How does IoT security and ZT convergence affect the interaction between users and technology}

\textbf{Both AL and GL remain largely silent on the human–technology interaction dimension of ZT and IoT security convergence}. Existing work focuses primarily on technical implementation rather than usability or user experience, as for example in the case of smart homes \cite{Dimitrakos2020} or wearable devices \cite{RTLACP}. However, ZT introduces continuous authentication, contextual access validation, and device trust scoring — mechanisms that fundamentally reshape how end users interact with connected systems.

These continuous verification processes can increase interaction friction for users, particularly in domains such as smart homes, healthcare, and wearables, where low-latency and seamless operation are essential. The literature on usable security \cite{cranor2008usableSecurity} and technology acceptance \cite{davis1989tam} indicates that excessive security prompts or cognitive load often lead to user fatigue, security bypassing, or decreased compliance. Integrating insights from these frameworks can help reconcile ZT’s security rigor with acceptable usability levels, supporting smoother adoption and compliance.

Future research should bridge Human-Computer Interaction (CI, \cite{Rogers2022}) and Usable Security (USEC, \cite{Garfinkel2014}) perspectives to design ZT mechanisms that are adaptive and minimally intrusive. For IoT contexts, lightweight, context-aware authentication and invisible trust re-evaluation could maintain ZT principles while preserving user experience and accessibility.

\subsubsection{RQ8: Benefits and costs of ZT  and IoT security convergence for individual users and society}

In AL, several works mention this topic for some specific communities or sectors of activity. GL does not address it; that might be explained by the fact that IoT security and ZT convergence GL papers are oriented towards organizations rather than end users or society. 

\textbf{Literature on this subject may be useful to end users and society}, as well as to organizations when they want to plan to realize IoT security and ZT convergence.   

The societal implications of ZT and IoT convergence extend beyond organizational boundaries. While ZT improves collective resilience by minimizing implicit trust and lateral movement, it also introduces increased energy and computational costs for IoT devices due to continuous encryption and verification. Moreover, the need for pervasive monitoring can amplify privacy risks and data-collection externalities. Few studies have quantified these trade-offs, yet cost–benefit modeling at the societal level is critical for guiding policy and regulation. Future research should incorporate sustainability and equity perspectives, examining how ZT–IoT convergence redistributes security costs and benefits across users, industries, and infrastructures. 

\subsubsection{RQ9: How does one balance user privacy and ZT and IoT security convergence requirements}
In AL, only \cite{Piya2021} surveys the adoption of IoT in healthcare, with an emphasis on ZT. In the GL considered, there is no discussion about that topic. \textbf{Privacy and the empowerment of users in the context of ZT and IoT security convergence are still open research fields}. 

Balancing ZT’s demand for pervasive telemetry with privacy principles remains a critical challenge. Continuous verification mechanisms depend on extensive device and user data, potentially conflicting with privacy-by-design and data minimization mandates. Emerging approaches—such as federated identity models, privacy-preserving authentication, and on-device trust evaluation—offer ways to maintain ZT compliance without excessive data centralization. Integrating guidance from the NIST Privacy Framework \cite{nistPrivacy2020} and aligning with GDPR \cite{GDPR2016} principles could help organizations achieve “privacy-aware ZT,” ensuring user trust in both technical and ethical dimensions of IoT ecosystems.

\section{Avenues for Future Research}
\label{S:discussion}

Divergence points from AL and GL are already identified in Section \ref{S:DataAnalysis} and \ref{S:results}, as well as under-explored research fields. Hereafter is a list of research gaps and opportunities deducted during the study. These gaps and opportunities are organized according to 3 different categories of research avenues identified during the study, requiring growing levels of interactions between communities: 
\begin{itemize}
    \item Yet-to-explore \textbf{technical under-explored research topics}, that are still open AL research topics. Such lists of topics can be found elsewhere in the existing literature, as it is usually identified from AL in both ZT/IoT security. This topics then does not require many interactions between industry and academics. This paper provide a list of subtopics identified during the review, with an emphasis on the topics not yet identified by the literature reviewed, as for example, attacks and forensic. 
    \item \textbf{Technical divergences} that requires to open up a dialogue in between academic security experts and industry to realign both research. Those topics were identified by searching for topics only covered by GL. For each of these research avenues, one has to determine why research transfer in between communities did not happen yet, and if there is any further research to do to apply AL research to industry. An example of such a technical divergence between AL and GL is identified in RQ1 (section \ref{RQ1}). It reflects a broader epistemic divide: academic work privileges formal security modeling and compliance to guidelines, while industry discourse is driven by deployability, regulatory timelines, and risk management constraints, thus advocating for IoT integration into ZTA. Bridging this gap requires hybrid research approaches — e.g., design-science studies, field experiments, and participatory action research — that can validate conceptual ZT models within operational IoT environments, and thus, may require collaborative work in between industry and academics.
    \item \textbf{Cooperative research topics}, that requires strong collaboration in between industry and research to be explored, due to their interdisciplinary and applied nature, as for example the benefit/cost estimate of the IoT security and ZT convergence. 
\end{itemize}

\subsection{Technical Under-Explored Research Topics}
\subsubsection*{Trust Interoperability}
Various mechanisms can be used in ZT to build trust in a device or subject, either with or without agents. IoT often requires to use  agentless trust evaluation because it is not possible to install it or make it run without consuming too many resources. 

It is then mandatory to use other means to build trust in IoT devices, the most studied one being either digital twin or reputation-based systems. Some of the AL reviewed were studying these methods (\cite{Jagannath2022} for example), but do not address the question of how to integrate these trust scores into a system-wide scheme. 

While this kind of trust score may be interoperable with system-wide trust score used in some products including system-wide trust and digital twins (for example \cite{paloaltowhitepaper} and  \cite{MicrosoftWhitePaper}), there is no formal method described to achieve this interoperability. 

\subsubsection*{Explaining Trust for IoT Devices}
Explainable AI received a lot of attention recently. While this topic is a buzzword in 2024, it is not identified in the related work given in section \ref{S:RelatedWork}. 

Explainable AI would enhance transparency by providing clear insights into how trust scores are determined, helping to identify and mitigate potential risks effectively. This transparency builds confidence among stakeholders by ensuring that trust decisions are based on understandable and justifiable criteria. Additionally, it facilitates compliance with regulatory requirements by offering auditable decision-making processes. 
While there is no paper on explainable AI, some discuss the implementation of Machine Learning techniques for various purposes. However, explaining security decisions made automatically is mandatory in an operational context; it would then be a research avenue for the future. 

\subsubsection*{Attacks} 

The corpus only contains 1 academic paper that focuses on attacks on platforms integrating IoT into a ZT architecture\cite{9781665410786}. This might be explained by the fact that ZT is still an active topic for defense and that most of the effort is now put into defining it properly rather than studying possible attacks and how to be resilient to those. 

However, ZT may be attacked using IoT as vectors, by using, for example, their number to organize a DDoS on PDP/PEP, similar to attacks on SDN controllers (see for example \cite{DBLP:journals/symmetry/AlashhabZADIA22}).

\subsubsection*{Forensic}
For ZT-based systems integrating IoT, forensic tools are critical for rapidly and reliably collecting, analyzing, and preserving immutable logs and device state data. How to adapt existing tools for Forensic IoT to be used in a larger ZT system is still an open research field. Specific forensic tools are essential for investigating in such systems because the architecture itself creates unique evidential challenges. ZT inherently limits access and mandates micro-segmentation, meaning evidence is highly distributed across numerous isolated devices and network zones, making traditional centralized collection ineffective.

\subsection{Technical Divergences}
\subsubsection*{Privacy Issues with ZT and IoT Security Convergence}
ZT and IoT security convergence presents privacy challenges. ZT often requires extensive data collection from devices to assess trust, especially when integrated into a ZT architecture, raising concerns about what data is gathered, how it is used, and who has access to it. ZT algorithms can be opaque, limiting user understanding of how data impacts their ability to interact with devices. Additionally, ZT might centralize vast amounts of user information from numerous devices, creating a potential target for misuse if not properly secured. To mitigate these concerns, organizations should focus on data minimization, anonymization, and providing users with clear explanations and control over their data privacy within the ZT framework.

While the privacy is addressed by some papers in AL, it is mainly focused on preserving information undisclosed, not using anonymization techniques but pseudo anonymization or de-identification, especially in some specific domains like healthcare \cite{Ali2021}, \cite{Chen2021}, \cite{Gofwen2023}, power IoT \cite{Huang2023}, \cite{Huang20232}, or 5G/6G \cite{YangYinghong2022}. It is then not focused on the end user privacy (except for \cite{RTLACP}), and more precisely on the end users control of their personal data.  

\subsubsection*{Blockchain-Based Convergence}
Blockchain-based convergence is an active topic in AL, but there is no mention of it in the GL. 

Blockchain is widely studied in AL for IoT. A lot of the interest in it is drawn from the decentralized nature of blockchain-based approaches. It well fits the problem of trusted interaction in between IoTs, and is compatible with the compliance of IoT security to ZT philosophy. However, IoT security integration to ZT relies on a relatively strong centralized control of security. As industry is more focused on the later approach, efforts should be put to realign the AL research on this topics with the industry approach, or to demonstrate the added value of blockchain-based approach in an industrial context.

\subsubsection*{Cryptography agility}
Cryptography is one of the topics that are addressed by AL but not GL. This may be explained by the traditional inertia of cryptograph, when it comes to adopt new standards, but also with the strong relationships between hardware and cryptography. Those two factors may slow the adoption of new techniques developed in AL by the GL, but existing work demonstrates that research must be undertaken to transfer work between communities. 

\subsubsection*{Trust Modeling}
While the convergence of ZT and IoT security presents a promising approach to enhancing the security posture of IoT systems, the role of trust modeling in this context remains largely under-explored in GL, while explored in AL. Further research is needed to develop and transfer to industry robust trust models that can effectively address the unique challenges posed by IoT environments, such as device heterogeneity, dynamic topologies, and the potential for adversarial attacks.

\subsection{Cooperative Research Topics}

\subsubsection*{Evaluation of Benefits and Costs of ZT and IoT Security Convergence}
As shown by answers of RQ questions related to \textit{measurement and values}, while the ZT and IoT security convergence is an active topic in both communities, there is little to no evaluation of benefits and costs of it, either from the industry, the society or the end user point of view. 

To be able to measure, evaluate and forecast such benefits and costs requires cooperation in between industry that has access to data, and experts from both security and social sciences to analyze, model and forecast benefits and costs. By now, there is no clear model for an organization planning to adopt such a convergence to estimate those values, that are mandatory to choose. 

\subsubsection*{User Acceptability}
While ongoing work on the convergence of ZT and IoT security offers significant potential for enhancing security, its acceptability by end user is unclear. Overly restrictive security measures can hinder usability and productivity, as demonstrated by the reluctance of users to adopt MFA, which is one of the core ZT technology. IoT that are in strong interactions with end-users (wearable devices for example) may reveal difficulties to converge. End-users may encounter challenges such as increased authentication steps, slower response times, or limited device functionality. To explore such a research field, it is mandatory to rely on industrial solutions and effective deployment, and to study with social sciences experts the impact of the convergence on end-users. 
\subsubsection*{Sustainable Cryptography}
ZT–IoT convergence demands cryptographic operations that are both resilient and energy-efficient. The challenge lies in maintaining continuous authentication and encryption across millions of constrained IoT nodes without compromising performance or sustainability. Cooperative research should focus on lightweight cryptography, post-quantum algorithm agility, and energy-aware key management. Integrating environmental metrics into cryptographic protocol evaluation would support the development of sustainable cybersecurity practices. Collaboration between cryptographers, hardware engineers, and environmental computing specialists could establish benchmarks for \textit{green cryptography}, balancing cryptographic robustness with reduced computational and energy overhead.

\subsubsection*{Governance and Policy Alignment}
The long-term success of ZT–IoT adoption depends on coherent governance frameworks that align technical design, regulatory compliance, and operational accountability. Existing initiatives—such as NIST’s ZT Architecture, ISO/IEC 27001 revisions, and the EU Cyber Resilience Act—offer complementary but fragmented guidance. Cooperative research should map these frameworks to identify overlaps and gaps, proposing harmonized compliance pathways for global interoperability. Engaging policymakers, industry consortia, and standardization bodies in joint studies will accelerate regulatory consistency, reduce implementation ambiguity, and enhance trust between cross-border IoT stakeholders.

\subsubsection*{In-Field Effective Convergence}
Deploying effective tools that seamlessly integrate ZT principles and IoT security presents a significant research challenge, demanding a collaborative effort between industry and academia, spanning both computer science and social science domains. This interdisciplinary approach is essential to address the complexities involved in implementing and adopting such solutions.

Researchers and industry practitioners must collaborate to devise innovative solutions for device authentication, secure communication protocols, and robust access control mechanisms, that may be acceptable by the users and still fit the requirements of ZT and the organizations infrastructure constraints.

Beyond technical challenges, the successful deployment of ZT and IoT security solutions hinges on user acceptance and adoption. Social scientists can contribute valuable insights into user behavior, privacy concerns, and the impact of security measures on usability. By understanding the human factors involved, researchers and developers can design more user-friendly and effective security solutions.

\subsubsection*{Human-Centric ZT}
Future research should prioritize the design of ZT architectures that integrate usability and human-behavioral considerations when ZT integrates IoT - thus increasing direct encounters between end users and ZT. Continuous verification and adaptive authentication mechanisms must be optimized for user experience, reducing cognitive and procedural friction. Collaboration between cybersecurity engineers, Human-Computer Interaction researchers, and behavioral scientists can help create models of usable ZT, combining strong access control with transparency and trust cues. Empirical studies—such as user-in-the-loop simulations and usability testing—will be critical to evaluate compliance and detect security fatigue, ensuring that security enforcement supports rather than hinders productive human interaction with IoT systems.
\section{Threats to Validity}
\label{S:threats}
Following \cite{Shadish2002}, we assessed construct, internal, external, and conclusion validity, as recommended in various previous studies \cite{10.1145/2915970.2916008}, \cite{Wohlin2000}.  Because our review is multivocal, we also assessed source credibility using the Garousi et al. \cite{GAROUSI2019101} checklist for grey literature (authority, methodology, objectivity, date/novelty, position vs. related sources, impact, outlet type). We give a summary of our checklist for quality assessment in appendix \ref{A:QA}, that are addressed by following a standard methodology. Below is a list of main threats to validity identified for this study. 

\subsubsection*{Construct Validity: database bias}
Most of the database used are oriented toward technical literature. This fact may have impacted by the validity of this study concerning less technical levels and activities. This bias may have been mitigated by including Google Scholar that gather publications for a broad number of disciplines

\subsubsection*{Construct Validity: search engine} Another bias may come from the search engine used when searching for GL. Recent studies, such as \cite{articleGoogle}, suggest that this bias for big companies comes from recent adjustments in Google algorithms, which leads to the disappearance of smaller website results compared to larger ones. We mitigated this by applying changes in the search we did, described in section \ref{S:A:searches}.

\section{Conclusion}
\label{S:conclusion}
This multivocal literature review offers the first consolidated analysis of how Zero Trust (ZT) and Internet of Things (IoT) security intersect across academic and industrial domains. This study identifies a pronounced epistemic divide: academic works predominantly pursue IoT compliance—rethinking device-level trust mechanisms and lightweight cryptography—whereas industrial grey literature promotes system-level integration within enterprise ZT architectures.

The study highlights four critical research gaps:
\begin{enumerate}
    \item Socio-technical factors—user acceptance, privacy implications, and organizational readiness remain underexplored.
    \item Measurement and validation—few studies empirically quantify the benefits or costs of IoT–ZT convergence.
\item Governance frameworks—there is limited discussion on migration paths and lifecycle management within organizations.
\item Standardization misalignment—industrial adoption largely follows NIST guidance, while academic research explores diverse conceptual models.
\end{enumerate}
Ultimately, the discussion underscores that ZT–IoT convergence is not a purely technical migration but a socio-technical transformation. Addressing it requires synchronized progress in standards, workforce development, privacy assurance, and public trust — the cornerstones of a resilient digital ecosystem.

Future work should establish joint evaluation frameworks between academia and industry, including longitudinal case studies, reference architectures, and cost–benefit analyses. Interdisciplinary collaboration—bridging computer science, human factors, and policy research—is essential for realizing ZT’s full potential in IoT ecosystems.

\section{Declarations}
\subsection{Data Availability}
As the paper is a Multivocal Literature Review, the data necessary to reproduce this work are mainly papers.

The paper contains references for both AL papers and GL artifacts. Papers from academic sources are available through the usual search engines and subject to various access restrictions, but most of them are freely accessible. 

Authors keep as an archive the industrial literature gathered during the MLR process. Many of the white papers considered were accessible only once provided personal information, and GL authors do not allow them to be freely redistributed. 

\begin{acks}Left blank for peer review.
\end{acks}

\bibliographystyle{ACM-Reference-Format}
\bibliography{ZT_IoT}


\appendix
\section{Detailed Methodology}
\label{S:A:methodology}

%

MLR considers two literature types: Academic Literature (AL) and Grey Literature (GL). AL consists of scholarly articles, conference papers, and books that have undergone rigorous peer review, ensuring a high level of credibility and methodological soundness. In contrast, GL includes non-peer-reviewed sources such as technical reports, white papers, government documents, and blog posts. While GL may lack the formal vetting process of academic publications, it often provides practical insights, recent developments, and real-world applications that are not yet captured in academic research. By synthesizing both AL and GL, an MLR leverages the strengths of each to address complex research questions more comprehensively and capture a wide range of perspectives and findings. 


GL is usually divided into 3 tiers depending on its reliability and quality \cite{Buck2021}:
\begin{itemize}
    \item High-quality GL (tier 1): This tier includes sources that are well-respected, often produced by reputable organizations such as government agencies, international bodies (e.g., WHO, UN), or large non-profit organizations. These documents often undergo some form of review or quality control and are cited frequently in both grey and academic literature. Examples include government reports, white papers from major research institutions, technical standards, and guidelines from reputable organizations.
    \item Medium-quality GL (tier 2): This tier includes sources that are generally reliable but may not have undergone rigorous peer review. They are typically produced by industry groups, smaller nonprofits, or other organizations with expertise in the field. While useful, these sources may contain some bias, particularly if they are produced by organizations with specific agendas or commercial interests. Examples include industry reports, conference proceedings, white papers from smaller organizations, and reports from think tanks. 
    \item Low-quality GL (tier 3): This tier includes sources that are less reliable, and often produced by individuals, small organizations, or entities without established credibility in the field. These documents may lack proper documentation, be anecdotal, or serve promotional purposes, making them less suitable for academic research. Examples include blog posts, opinion pieces, newsletters, and informal publications from lesser-known sources.
\end{itemize}
This work is focused on tiers 1 and 2, as the results of the searches gave a few results belonging to tier 1 (5 out of 36). Tier 1 papers are for example NIST SP 800-207 \textit{Zero Trust Architecture} \cite{NIST} ; Tier 2 papers are mainly industrial white papers, such as \textit{Zero Trust Cybersecurity for the Internet of Things} \cite{MicrosoftWhitePaper} from Microsoft.

The MLR was conducted based on the process defined by Garousi et al \cite{GarousiFM17}, while following PRISMA 2020 framework guidelines \cite{Page2021PRISMA} for systematic reviews. Garousi's process is divided in phases: Planning phase, conducting phase, and reporting phase. In the planning phase, one has to establish the need for research, define its goal, and formulate research questions to guide the investigation. Then, the conduct phase consists of literature searches via databases and grey literature via web search. During that phase, one also has to define inclusion and exclusion criteria for paper selection. Finally, the reviewing phase consists of systematically extracting and synthesizing data from selected studies and deriving meaningful insights.

\section{Search Strategies}
\label{S:A:searches}
\subsection{Database Search for AL}
Key terms related to the search are \textit{zero trust}, and \textit{IoT}, and serve as the basis for constructing search queries. 
The following (case-insensitive) search string was defined:

\begin{quote}
("zt" AND "iot") OR ("zt" AND "internet of things") OR ("zero trust" AND "iot") OR ("zero trust" AND "internet of things") 
\end{quote}

The following academic databases were used: IEEE Xplore, ACM Digital Library, Science Direct, and Google Scholar. It has been decided to consider only documents published between January 2014 and November 2023. The title and abstract search yielded 93 articles without duplicates.

 After that, data has been collected to determine the papers suitability for inclusion in the study based on inclusion and exclusion criteria. Thereafter, titles, abstracts, and keywords were screened to assess the relevance and appropriateness of terms used to index articles, and include only literature closely aligned with the research focus for further analysis. The study only included items that were published in peer-reviewed journals and conferences, books, and thesis reports that are ranked on CORE \cite{core} or SCIMAGO \cite{scimago}.
Exclusion criteria were (1) items that are preprints, graduate projects, master's and technical reports
and (2) articles that don't include ZT and IoT.

Finally, backward snowballing has been performed. At first, as an automated process by using Buhos \cite{BUSTOSNAVARRETE2018360}, a web-based tool that manages the systematic literature review. 
Automated snowballing helped retrieve one new document. A manual snowballing was also performed, as some references were not parsed correctly by the tool, as an alternative to automated review. The papers in the reference lists were evaluated using the inclusion and exclusion criteria. 3 additional AL papers were retrieved this way.

The structured literature search generated 68 AL articles. 

\subsection{Web Search for Grey Literature}

The study relied on the following search in Google for documents written in English matching the following search string in their title, using the same methodology to construct a search string for GL in MLR as \cite{GAROUSI2019101}, \cite{DBLP:journals/infsof/GarousiFM19} or \cite{Buck2021}: 
\begin{quote}
("zt" AND "iot") OR ("zt" AND "internet of things") OR ("zero trust" AND "iot") OR ("zero trust" AND "internet of things") 
\end{quote}
The query was split in independent subqueries, and results aggregated after this first search. This gave a total of 262 documents. However, the number of valuable results was very low: out of those 262 documents: (1) 34 were duplicates of the AL review, (2) 175 were blogs/news/event announcements, most of them coming from websites of major industrial actors, and (3) 40 videos or podcasts. At last, only 13 white papers dealing with the topic remained, most of them coming from major actors (such as Microsoft, Palo Alto, Fortinet, ...), 1 being provided by a public organization (IEEE), while all the other came from software providers. It is unclear to us if those numbers come from the topic or a shift in the way the industry communicates. Indeed many technical blog notes and videos that might be of interest were found, but are not considered as relevant in the usual MLR methodology .  

Another bias may come from the search engine used. Recent studies, such as \cite{articleGoogle}, suggest that this bias for big companies comes from recent adjustments in Google algorithms, which leads to the disappearance of smaller website results compared to larger ones. A modified  query was then used, specifically focusing on white papers written in English and that should contain text matching the following query:

\begin{quote}
"white paper" AND (("zt" AND "iot") OR ("zt" AND "internet of things") OR ("zero trust" AND "iot") OR ("zero trust" AND "internet of things"))
\end{quote}

This gave a total of 163 documents, retrieved between November 2023 and March 2024\footnote{The retrieval process was long due to the fact that entreprise often requires you to provide business contact and other details before being able to retrieve documents.}. Out of those, 108 white papers were accessed, most of the time not directly, but by browsing through the websites and giving business details to retrieve the document. Removing duplicates (found either using this search or the previous one) gave a total of 93 white papers, 16 coming from government agencies, 6 coming from industrial consortia, and 72 remaining from the industrial sector. 

The quality and relevance to the topic of the papers were assessed, and lead to the exclusion of 55 of them. A snowballing pass to retrieve relevant papers cited in GL was performed, but no additional papers was found, leading to 36 GL papers to review.

\section{Data Analysis: Topic Description}
\label{S:AppendixDataAnalysis}
\begin{itemize}
    \item \textbf{Models}: papers discussing the ZT model. Both GL and AL papers discuss the ZT model and how to implement it in the IoT context. However, the way it is discussed in AL and GL differs. In AL, papers that discuss this topic present the model, the pros and cons of it, and how it should be modified.  In GL, papers usually compare segmentation-based and ZT to show the pros of integrating IoT into the perimeter of a ZT architecture, but do not propose to modify the model, and usually comply with the NIST ZT standard architecture. 
    \item \textbf{Access control}: papers discussing access control paradigm for IoT (ABAC, dynamic access control, ...). Many AL papers discuss new access control paradigms designed to respect the ZT pillars for IoT. GL papers falling in this category discuss the centralized management of access control and the tools to do it. 
    \item \textbf{Authentication and Identification}: papers about authentication methods, continuous authentication, decentralized authentication and identification, MFA for IoT. AL papers discuss decentralized authentication and new methods to identify/authenticate devices or subjects using devices. On the other hand, GL discusses MFA for IoT, the use of a gateway to provide strong authentication to brownfield devices, for example. 
    \item \textbf{Privacy}: AL papers often discuss privacy. The review conducted respect AL authors' choices and count papers dealing with this topic when one of the keywords they chose was privacy. None but one of the papers is centered on privacy issues and their impact. 13 out of 36 GL papers discuss about privacy briefly, mainly to list it as a threat. Papers in GL dealing with healthcare contain a slightly longer description of privacy threats, but without any formal analysis of them. 
    \item \textbf{Cryptography}: Adoption of new cryptographic protocols or ideas is usually done very slowly compared to the pace of adoption of new security models or ideas. That may explain why papers discussing cryptography are mainly AL ones. 
    \item \textbf{Applications}: papers discussing application domains. AL papers mainly discuss implementation challenges and/or provide feedback on adopting ZT for specific domains that use IoT, such as healthcare, IIoT, or smart home. Some GL papers present specific implementations of ZT and IoT security convergence for healthcare and IIoT. 
    \item \textbf{Infrastructure}: fog/edge/MEC and other device-to-device infrastructure underlying the IoT deployment, and the specificity of this context for ZT. AL and GL papers present an adaptation of the ZT model for those platforms, a specific implementation for edge servers, for example. 
    \item \textbf{Blockchain}:  this study shows that while there is a lot of work published in AL concerning blockchain-based authentication, this is not a choice being documented in GL, as just one of the white papers reviewed was discussing blockchain (an IEEE standard for blockchain-based ZT framework for the IoT \cite{IEEE}).  
    \item \textbf{Tools}: papers discussing tools dedicated to ZT and IoT security convergence, such as Digital Twin and reputation systems to estimate whether a device is compromised, IDS, firewalls, and so on. Both GL and AL papers present tools that are implemented to be deployed specifically in the context of ZT and IoT security convergence. 
    \item \textbf{Power IoT}: Power Grid IoT, i.e., IoT deployed on power grids.  The review found only AL papers discussing this topic, all of them coming from China. 
    \item \textbf{Trust}: trust score, trust evaluation, management, and modeling for IoT. Mainly discussed in AL, those aspects are just often mentioned in GL, without further description of them. More details can be found in the 3-tier GL; the way those subjects are mentioned in the 2-tier GL possibly indicates that those subjects are confidential to the industry. 
    \item \textbf{IIoT}: Both AL and GL discuss how to apply, implement, and/or modify ZT model in this context.
    \item \textbf{Network}: Network aspects of ZT and IoT security convergence. Topics covered in both AL and GL include the deployment point of view, network infrastructure, and specificity on policy enforcement in an IoT environment. 
    \item \textbf{AI}: Artificial intelligence applied to ZT and IoT security convergence. Papers in AL and GL discuss digital twins, trust evaluation, and intrusion detection. 
    \item \textbf{5G/6G}: ZT and IoT security convergence in the context of 5G/6G. Both AL and GL discuss how ZT can be implemented and/or modified for use in this context. 
    \item \textbf{Attacks}: attacks in the context of ZT and IoT security convergence. Just a few papers discuss attacks in such a context, mainly from defense point of view. 
\end{itemize} 

\section{Results: detailed RQ/Literature Mapping}
\label{S:AppendixMapping}
\subsection{Level of Analysis: Concept and Architecture}
\subsubsection{RQ1: Technical convergence of IoT security and ZT}
One may differentiate work dealing with IoT compliance to ZT philosophy from those oriented toward integration into an organization-wide ZT system. 

\textbf{Concerning AL, on the IoT compliance with ZT philosophy side}, the most considered topic is (continuous) authentication and identification. \cite{Alshomrani2022} presents continuous authentication techniques for IoT devices using physical functions to integrate an IoT device seamlessly into ZT. \cite{Shah2021} also provides continuous authentication techniques, but is oriented toward device-to-device authentication. \cite{Meng2022} proposes a blockchain-based protocol to remove the trust authority involved in mutual authentication between IoT, arguing that this trust authority is granted an implicit trust. \cite{Zhao2021} presents a method to authenticate users using blockchains in an IoT environment. \cite{Liu20232} introduces lightweight protocols to check identity on IoT. \cite{Lei2023} proposes to use physical-layer security to enhance ZT in IIoT, including authentication, but the paper has a larger scope because it merges the physical-layer security within the ZT model. \cite{Zawadzki2022} discusses how a smart manufacturing system may comply with ZT.
 
Access control is addressed by 4 papers: \cite{RTLACP} introduces a lightweight access-control protocol for wearable devices. \cite{Dhar2021} and \cite{Li2023} introduce blockchain-based methods to assess a trust score for IoT and control access. \cite{Awan2023} proposes a blockchain-based attribute-based access control (ABAC) for IoT.

Data protection is a subject addressed by two papers: \cite{Han2022} and \cite{Wang20232} provide a ZT blockchain-based data storage scheme, where data are stored on IoT. 

Other work enforcing the compliance of IoT to ZT standard includes \cite{Lee2021} that implements lightweight Software Defined Perimeter (SDP) at the IoT level.

\textbf{Concerning AL, on the integration of IoT in a ZT architecture}, some works deal with trust: \cite{Abuhasel2023} introduces an algorithm evaluating the risk of access to an IIoT device, that can be deployed in a TE.  \cite{Alawneh2022} makes the bridge between the notion of trust computing used in IoT and how it may be integrated into ZT. \cite{Jagannath2022} provides a framework to integrate Digital Twins (DT) of IoT into ZT architecture to provide a trust score to ZT. \cite{Munasinghe2023} presents a method to calculate the Trust Score using AI. \cite{Liu2023} proposes a blockchain-based solution to integrate IoT devices in ZT while preserving anonymity in a cryptocurrency environment. \cite{YangYinghong2022} does the same without targeting a specific environment.

Dealing with modifying, enhancing, or adapting the ZT principles to IoT, \cite{Ameer2022} and \cite{Ameer2023} propose to enhance ZT with an authorization framework based on the score to deal with uncertainty on whether the device is compromised and its identity.
\cite{Chen2021} presents how to leverage ZT to 5G smart healthcare platforms, which implies considering additional entities when evaluating trust, as 5G runs on a shared infrastructure. \cite{Hao2021} presents a ZT blockchain-based information-sharing protocol between IoT; this work, however, breaks the assumption of centralized management of trust between entities. \cite{Kailash2023} proposes a framework to integrate IoT into ZT, oriented toward using gateways, similar to the Azure approach. \cite{Kobayashi2023} studies the integration of CPS into ZT architecture and proposes a framework to deal with such platforms. \cite{Palmo2023} presents a framework to integrate IoT in SDP. \cite{Reaz2022} presents a protocol to onboard new IoT devices into a system without fully trusting its supply chain. \cite{Srivastava2023} introduces a framework containing actual guidelines and recommendations to deploy ZT in an IoT environment, with details on how to do it technically. \cite{Szymanski2022} provides a framework that includes ZT and IoT to deal with the post-quantum era. \cite{Zhang2023} proposes a method to implement ZT in the context of an IoT data security management system. 

GL mainly advocates integration rather than making the IoT more compliant with ZT philosophy. Some literature deals with manufacturing ZT-compliant IoT. However, the vast majority of the GL explains how to integrate IoT into the ZT system.

 \subsubsection{RQ2: Technical benefits and limitations of IoT security and ZT convergence}
 In both AL and GL, the question of the benefits and limitations of IoT security and ZT convergence is rarely and briefly discussed. It is more often implicit that the choice of the architecture shows if the choice has been made to isolate IoT resources or to embed them in the system policies.

 In AL, one paper discusses the opportunity of integrating IoT in ZT systems and decides to let them apart. On the other hand, some papers define strategies to include them, but without discussing if including IoT is beneficial. \cite{Samaniego2018} states that IoT cannot be trusted, and thus provides a blockchain-based framework for managing (and thus isolating from the system) IoT devices. \cite{Uehara2021} provides a framework to isolate IoT in a mist architecture, and, by doing so, avoid the complexity of IoT integration. 

 In GL, some papers explicitly give integration techniques, for example, by the use of gateway or micro-segmentation, while others isolate the devices.
 
 \subsubsection{RQ3: Human resources needed to maintain a system that adopted a ZT and IoT security convergence philosophy}

Shifting from traditional security to ZT is a process requiring expertise in both ZT and the targeted systems \cite{ZTNA}. IoT security and ZT convergence add the need for IoT skills and a good knowledge of operational conditions of exposure to the risk of any of the IoT considered.  

AL does not address this topic. GL does not either, while some offers include training sessions; it is nevertheless a key point in the ZT and IoT security convergence. 

 
 \subsection{Level of analysis: Firms and Industries}
 \subsubsection{RQ4: How can an organization realize IoT security and ZT convergence} 

 How to realize IoT security and ZT convergence is explored through the angle of sectors of activity in AL.  

 \cite{Ali2021} introduces techniques to gather information in healthcare systems, including IoT to implement ZT.
\cite{Gao2021} and \cite{Wang2023} present techniques for implementing ZT into Power IoT. 
\cite{Chen20212} provides blockchain-based technologies to secure data flows into a power network system. \cite{terminals} presents methods to decentralize the trust score computation in the same context, using Federated Learning and blockchains. 
\cite{Huang2023} and \cite{Huang20232} introduce a method to continuously monitor and provide ABAC, and integrate it into a TE for power IoT. \cite{Dimitrakos2020} does the same in the context of a smart home.
\cite{Federici2023} discusses how to implement ZT for IIoT ; so does \cite{Zhang2021}, with a specific focus on continuous authentication. \cite{Kobayashi2023} studies the integration of CPS into ZT architecture and proposes a framework to deal with such platforms. 
\cite{Nour2023} introduces an ML-supervised IDS to integrate in a ZT-based 5G environment. \cite{Sedjelmaci2023} also introduces an IDS to integrate into a ZT-based system, but targets 6G instead of 5G.
\cite{Kondaveety2022} introduces a ZT framework for connected vehicles. \cite{Li20232} presents methods that focus on the security of edges and how to integrate them into ZT architectures. \cite{9781665410786} introduces a method to prevent APT attacks in an IoT based on ZT model. \cite{Tyler2021} provides the results of experiences in using a ZT framework and following Cisco guidelines in a healthcare environment. 

In GL a lot of papers discuss how to realize ZT and IoT security convergence, without specifying for which domain it may be relevant to do this convergence. GL only discusses the specificity of two sectors: healthcare and IIoT.

 \subsubsection{RQ5: How does IoT security and ZT convergence provide added value for organizations}
 
 In AL there is little to no attempt to measure the added value for organizations about this convergence. In this corpus, only \cite{marketing} compares perimeter-based security and ZT in the context of IIoT. 

 In GL, this topic is massively addressed, as it is a key factor in the adoption of this convergence in the organization. Indeed, the promise of stronger security is in most of the GL reviewed. Demonstrating the added value of ZT and IoT security convergence is done by comparing it to segmentation-based security. 
 The drawbacks of such a convergence, often listed when dealing with ZT adoption (see for example \cite{ZTNA}), are not addressed in GL.

 \subsubsection{RQ6: Organizations setup, government and development of IoT security and ZT convergence}
One may foresee the complexity of the process induced by a shift from an existing IoT security to the convergence of IoT and ZT by looking at the complexity of shifting from perimeter-based security to ZT. Shifting from perimeter-based security to ZT is a long process that usually lasts for years, carefully migrating segments one after the other, with meticulous transcription of policies \cite{ZTNA}. 

The review did not find either AL or GL dealing with the subject of how an organization may manage these projects. Two papers in this study discuss factors influencing the organization of this process, focusing on healthcare systems. \cite{Samah2023} studies factors behind the adoption of ZT and IoT in the Malaysian healthcare system. \cite{Gofwen2023} analyses the impact on privacy of IoT integration in ZT, and the impact on the adoption of IoT in healthcare systems.
 
 \subsection{Level of Analysis: Users and Society}
There are very few papers, either in AL or GL, that consider ZT and IoT from users and society level of analysis. It could be either an artifact of the methodology\footnote{The literature considered, as the search methods are oriented toward research (so, technical) papers.}, or a real lack of support for those aspects.  
 
\subsubsection{RQ7: How does IoT security and ZT convergence affect the interaction between users and technology}
There is no paper, neither in AL nor GL, that addresses this issue. While the review found papers about IoT security and ZT convergence in the context of smart homes \cite{Dimitrakos2020} or wearable devices \cite{RTLACP}, both of them study the technical point of view and not the adoption by end users.

\subsubsection{RQ8: Benefits and costs of ZT  and IoT security convergence for individual users and society}

In AL, several works mention this topic for some specific communities or sectors of activity. As stated before, \cite{Samah2023} studies factors behind the adoption of ZT and IoT in the Malaysian healthcare system. \cite{Gofwen2023} analyses the impact on privacy of IoT integration in ZT, and the impact of the adoption of IoT in healthcare systems. The primary focus of those works is the impact on the organization and not end users. GL does not address this topic; that might be explained by the fact that IoT security and ZT convergence GL papers are oriented towards organizations rather than end users or society.

\subsubsection{RQ9: How does one balance user privacy and ZT and IoT security convergence requirements}
In AL, only \cite{Piya2021} surveys the adoption of IoT in healthcare, with an emphasis on ZT. The review did not find any discussion about that topic in GL. Privacy and the empowerment of users are still open research fields. 

\section{GL document list}
\label{A:GLdoclist}

\small
\begin{longtable}{p{4.0cm}  p{3.0cm}  p{2.3cm}  p{5.0cm}}
\caption{GL Artifacts Summaries}\\
\toprule
\textbf{Title} & \textbf{Enterprise} & \textbf{Year} & \textbf{Summary} \\ \midrule \endfirsthead
\multicolumn{4}{c}{{\bfseries \tablename\ \thetable{} -- continued from previous page}} \\
\hline
\textbf{Title} & \textbf{Enterprise} & \textbf{Date} & \textbf{Summary} \\ \hline \endhead
\hline \multicolumn{4}{r}{{Continued on next page}} \\ \endfoot
\bottomrule \endlastfoot
5 Must-Haves for Comprehensive Zero Trust  IoT Security  & Palo Alto Networks, Inc & 2023 & List of tools required to achieve IoT and ZT integration\\
\hline
5 Steps to  Zero Trust for  Unmanaged  and IoT Devices   & ordr & 2024 & Recommendations for isolation and micro-segmentation of IoT/unmanaged devices (white paper)\\
\hline
Adopting Zero Trust Security for Healthcare: 3 Main Drivers that Make the Case & Cynerio & 2023 & Recommendations for ZT in healthcare with IoT support (white paper)\\
\hline
Achieving  Zero Trust for Connected Devices
 & netskope & 2023 & Recommendations for IoT integration in ZT framework (white paper)\\
\hline
Architecting Security into Your 
Modern Enterprise & epam & 2020 & Recommendations for IoT integration in ZT framework (white paper) \\
\hline
Architecting the Zero 
Trust Enterprise & Palo Alto Networks, Inc. & 2021  & General recommendations for ZT architecture (white paper)\\
\hline
Arista Zero Trust Security for Cloud Networking  & Arista & 2023 & General recommendations for ZT architecture (white paper) \\
 \hline
Bolstering Enterprise  Security Using  Zero Trust  Architecture  & cradlepoint (part of Ericsson) & 2023  &  General recommendations for ZT architecture (white paper) \\
\hline
Total Visibility:
The Master Key to Zero Trust Security & Forescout & 2021  & General recommendations for ZT adoption (white paper)  \\
 \hline
Zero Trust
Manufacturing & KEYFACTOR & 2022 &  Recommendations to implement ZT in a system where IoT comes from complex supply chains (white paper)\\
\hline
Best Practices for Extending Zero Trust to  Government Networks & Forescout & 2021 & Recommendations for US federal agency's adoption of ZT with a strategy for handling IoT (white paper)\\
 \hline
A Unified Architecture for 
Achieving Zero Trust Across 
all Network Domain & EMA &  2021 &  Recommendations for ZT architecture across networks domains (white paper)\\
\hline
MEC security;
Status of standards support 
and future evolutions & ETSI & 2022 &  Recommendations for MEC security based on ZT (white paper) \\
 \hline
Extending Zero-Trust Security 
to Industrial Operations & Cisco & 2021 &  Recommendations for  ZT in IIoT (white paper) \\
\hline
Fijoport Zero Trust Architecture & Fijowave Cyber Security & 2022 & Description of a solution for IoT security based on ZT\\
\hline
Edge Security Essentials & Dell Technologies & 2024 &  Recommendations for  ZT at edges (white paper) \\
\hline
Identity and Zero Trust: 
a HEALTH-ISAC guide for CISOS  & H-ISAC & 2024 & Recommendations for merging ZT and HEALTH-ISAC framework identities management\\
\hline
Zero Trust Manufacturing  & Farallon Technology Group \& KEYFACTOR & 2021 & Challenges and best practices to implement ZT in a system where IoT comes from complex supply chains (white paper)\\
\hline
Implementing zero 
trust for industrial 
environments & Claroty (ITSARI) & 2022 & Recommendations for IoT integration in ZT framework (white paper)\\
\hline
IoT Security: Choose a flexible zero trust approach to secure 
nontraditional devices in your digital terrain & Forescout & 2022 &  Recommendations for IoT integration in ZT framework\\
\hline
Front-End Access Control (FEAC) & SafePay Systems & 2021 & ZT Access Control for IoT (deliverable from  European Union’s Horizon 2020 IoTAC) \\
\hline
Moving To An Endpoint-Centric Zero Trust Security Model with SentinelOne & SentinelOne & 2021 &  Recommendations for moving to ZT framework an IoT environment (white paper)\\
\hline
Security:  Creating Trust in a Zero Trust World & opentext & 2021  & Recommendations for adopting ZT framework (white paper)\\
\hline
The Right Approach to Zero Trust for Medical  IoT Devices & Palo Alto Networks, Inc & 2023 & Recommendations for implementing ZT framework for medical IoT (white paper)\\
Secure IoT begins with Zero-Touch Provisioning at scale & infineon & 2021 & Recommendations for IoT identity management for ZT\\
\hline
Securing a New Digital World with Zero Trust: How Zero Trust Cybersecurity is Transforming the IoT Industry
 & HIKVISION & 2021 & Recommendations for deploying ZT on IoT for video surveillance (white paper)\\
\hline
Seven Elements of Highly Successful  Zero Trust Architecture & Zscaler  & 2022 & Recommendations for ZT architecture implementation \\
\hline
What is Clinical Zero Trust? & MEDIGATE & 2024  & Recommendations for implementing ZT framework for medical IoT in a clinical environment (white paper)\\
\hline
Embracing a Zero Trust hardware access security model & Sepio & 2024 & Recommendations of how to enforcing endpoint an IoT access in ZT (whitepaper) \\
\hline
Zero Trust Edge Computing & mainsail & 2023 & Recommendations of how to leverage Western Digital Ultrastar Edge, Red Hat Enterprise, and Mainsail Metalvisor to implement ZT at the edges (whitepaper) \\
\hline
SD-WAN for  Manufacturing & Aruba & 2022 & How SD-WAN can be part of a SASE ZT framework to secure IoT devices (technical paper)\\
\hline
Securing Digital Innovation 
Demands Zero-trust Access & Fortinet & 2021 & Recommendations for adoption of ZT at network edge (whitepaper) \\
\hline
Closing hidden security 
gaps in zero trust 
architectures:
what public sector needs to know & Armis, Inc. & 2022 & Public sector oriented to introduce ZT and specific challenges, including unmanaged assets and IoT (white paper)\\
\hline
Zero Trust Cybersecurity for the  Internet of Things & Microsoft Azure & 2021 & Introduction to Azure solution to ZT+IoT convergence (white paper)\\
\hline
Accelerate Zero Trust Adoption  & ExtraHop & 2022 & Recommendations to accelerate ZT adoption (white paper) \\
\hline
Zero Trust 
Architecture for 
advancing mobile 
network security 
operations
 & Ericsson & 2024 & Introduction to ZTA convergence with mobile networks (white paper)\\
\hline
IoT and IIoT security: Achieving Zero Trust with Secure Device Identities & NEXUS, IN groupe & 2024  & Identity management for IoT and IIoT in a ZT environment (white paper)\\
\hline
 
\end{longtable}

\section{Quality Assessment Checklist}
\label{A:QA}
\begin{table*}[ht]
\centering
\caption{Quality and validity assessment criteria for included studies in the MLR}
\label{tab:validity-coding}
\renewcommand{\arraystretch}{1.3}
\begin{tabular}{|p{7cm}|p{7cm}|}
\toprule
\textbf{Indicators} & \textbf{Solution provided in this MLR}  \\
\midrule
\rowcolor[gray]{0.9}
\multicolumn{2}{|c|}{\textbf{Construct Validity}} \\
\hline
\multicolumn{2}{|p{14cm}|}{
Extent to which the study correctly operationalizes and measures its theoretical concepts. 
Evaluate whether key constructs (e.g., “IoT”, “Zero Trust”) are clearly defined and linked to measures.
}\\
\hline
• Clear definition of constructs \newline
• Alignment between theory and measurement \newline
• Use of established definitions or metrics & 
\checkmark extensive literature about concepts \newline 
\checkmark results in line with literature \newline 
\checkmark standard methods applied \\ 
\hline
\rowcolor[gray]{0.9}
\multicolumn{2}{|c|}{\textbf{Internal Validity}} \\
\hline
\multicolumn{2}{|p{14cm}|}{
Confidence that study results are not affected by confounding variables or design bias. 
Focus on rigor of design, control of bias, and reasoning chain.
}\\
\hline
• Methodological transparency \newline
• Bias/limitations discussed \newline
• Triangulation or validation methods &
\checkmark Detailed methodology provided \newline 
\checkmark Dedicated section \newline 
\checkmark Different data sources, investigator triangulation \\ 
\hline
\rowcolor[gray]{0.9}
\multicolumn{2}{|c|}{\textbf{Conclusion Validity}} \\
\hline
\multicolumn{2}{|p{14cm}|}{
Whether conclusions are justified given the evidence presented. 
Assess logical coherence and whether conclusions overstate findings.
}\\
\hline
• Data support claims \newline
• Discussion of uncertainty \newline
• Statistical/qualitative justification & 
\checkmark Detailed mapping in appendix \newline 
\checkmark Discussed  \newline 
\checkmark Provided in tables and figures \\
\hline
\rowcolor[gray]{0.9}
\multicolumn{2}{|c|}{\textbf{External Validity}} \\
\hline
\multicolumn{2}{|p{14cm}|}{
Extent to which findings can be generalized to other contexts, populations, or applications.
}\\
\hline
• Context description \newline
• Discussion of generalizability \newline
• Replicability of setting or sample &
\checkmark Detailed description given \newline 
\checkmark Not applicable  \newline 
\checkmark Full list of selected artifact provided \\
\hline
\rowcolor[gray]{0.9}
\multicolumn{2}{|c|}{\textbf{Source Credibility (MLR}} \\
\hline
\multicolumn{2}{|p{14cm}|}{
Specific to multivocal reviews: credibility and trustworthiness of each source, based on Garousi et al. (2019) guidelines for grey literature.}\\
\hline
• Tiers classification of GL \newline
& \checkmark Used \\
\hline
\end{tabular}
\end{table*}

\end{document}